\documentstyle[12pt]{article}
\newcommand{\ab}[2]{\begin{array}{c}#1\\#2\\~\end{array}}
\newcommand{\ad}{\mbox{ad}}
\newcommand{\ben}{\begin{equation}}
\newcommand{\een}{\end{equation}}
\newcommand{\bea}{\begin{eqnarray}}
\newcommand{\eea}{\end{eqnarray}}
\newcommand{\nn}{\nonumber \\ }
\newcommand{\hf}{\frac{1}{2}}
\newcommand{\th}{\frac{1}{3}}
\newcommand{\thf}{\frac{3}{2}}
\newcommand{\im}{\mbox{im }}
\newcommand{\ke}{\mbox{ker }}
\newcommand{\pa}{\partial}
\newcommand{\A}{\alpha}
\newcommand{\B}{\beta}
\newcommand{\D}{\delta}

\newcommand{\G}{\gamma}

\newcommand{\LM}{\Lambda}

\newcommand{\OM}{\Omega}

\newcommand{\ca}{\mbox{$\cal{A}$}}
\newcommand{\cg}{\mbox{$\cal{G}$}}
\newcommand{\ch}{\mbox{$\cal{H}$}}

\newcommand{\cw}{\mbox{$\cal{W}$}}

\newcommand{\C}{\mbox{\rm\hspace{.0em}\rule{0.042em}{.674em}
\hspace{-.642em} C$\,$}}

\newcommand{\Z}{\mbox{$Z$\hspace{-1.1em}$Z$ }}
\newcommand{\N}{I\hspace{-0.27em}N}
\newcommand{\z}{\mbox{\footnotesize Z\hspace{-1mm}Z}}
\newcommand{\1}{\mbox{\hspace{.42mm}\rule{0.2mm}{2.8mm}\hspace{-2.24mm}
{\large 1}}}
\def\ord#1#2{{#2\over (z-w)^{#1}}}
\def\ordo#1{{#1\over z-w}}
\def\ope#1#2{#1(z)#2(w)}
\newcommand{\NPB}[1]{{\it Nucl. Phys.} {\bf B#1}}
\newcommand{\PLB}[1]{{\it Phys. Lett.} {\bf B#1}}
\newcommand{\MPLA}[1]{{\it Mod. Phys. Lett.} {\bf A#1}}
\newcommand{\IJMPA}[1]{{\it Int. J. Mod. Phys.} {\bf A#1}}
\newcommand{\CMP}[1]{{\it Comm. Math. Phys.} {\bf #1}}

\newcommand{\norm}[1]{{\protect\normalsize{#1}}}
\newcommand{\LAP}
{{\small E}\norm{N}{\large S}{\Large L}{\large A}\norm{P}{\small P}}
\newcommand{\be}{\begin{equation}}
\newcommand{\ee}{\end{equation}}
\newcommand{\ena}{\end{eqnarray}}
\newcommand{\beano}{\begin{eqnarray*}}
\newcommand{\enano}{\end{eqnarray*}}
\newcommand{\sect}[1]{\setcounter{equation}{0}\section{#1}}

\newcommand{\hs}[1]{\hspace{#1 mm}}

\newcommand{\half}{\frac{1}{2}}

\newcommand{\cb}{\mbox{$\cal{B}$}}

\newcommand{\eps}{\epsilon}

\newcommand{\wh}[1]{\widehat{#1}}
\newcommand{\mb}[1]{\hs{5}\mbox{#1}\hs{5}}

\newtheorem{prop}{Property}

\newtheorem{theo}{Theorem}

\newtheorem{lem}{Lemma}

\newcommand{\PL}[1]{Phys.\ Lett.\ {\bf #1}}

\newcommand{\IJMP}[1]{Int. Journ.\ Mod.\ Phys.\ {\bf #1}}

\begin{document}
\begin{titlepage}
\null
\hbox to \hsize{\vbox{\hsize 8 em
{\bf \centerline{Groupe d'Annecy}
\ \par
\centerline{Laboratoire}
\centerline{d'Annecy-le-Vieux de}
\centerline{Physique des Particules}}}
\hfill
\hfill
\vbox{\hsize 7 em
{\bf \centerline{Groupe de Lyon}
\ \par
\centerline{Ecole Normale}
\centerline{Sup\'erieure de Lyon}}}} 
\hrule height.42mm

\vspace{.42cm}
\begin{center}
  {\Large\bf Secondary Quantum Hamiltonian Reductions \par}
  \vspace{.7cm}
  \baselineskip=7mm

  {\large Jens Ole Madsen\footnote{\noindent email : madsen@lapphp8.in2p3.fr}
and
Eric Ragoucy\footnote{\noindent email : ragoucy@lapp.in2p3.fr} \par}

\vspace{4mm}

{\sl Laboratoire de Physique Th\'eorique\footnote{URA 1436 du CNRS 
associ\'ee \`a l'ENS de Lyon et \`a l'Universit\'e de Savoie} \LAP, \\
groupe d'Annecy, \\
LAPP, Chemin de Bellevue, B.P. 110, \\
F-74941 Annecy-le-vieux Cedex, France. \par}

\vfill

{\bf Abstract}
\end{center}
\par

Recently, it has been shown how to perform the quantum hamiltonian reduction in
the case of general $s\ell(2)$ embeddings into Lie (super)algebras, and in the
case of general $osp(1|2)$ embeddings into Lie superalgebras. 
In another development it has been shown that when \ch~and $\ch'$ are both
subalgebras of a Lie algebra \cg~with $\ch'\subset\ch$, then classically the 
$\cw(\cg,\ch)$ algebra can be obtained by performing a secondary
hamiltonian reduction on $\cw(\cg,\ch')$. 
In this paper we show that the corresponding statement is true also for quantum
hamiltonian reduction when the simple roots of
$\ch'$ can be chosen as a subset of the simple roots of \ch. 

As an application, we show that the quantum secondary reductions provide a 
natural framework to study and explain the linearization of the \cw\ algebras,
as well as a great number of new realizations of \cw\ algebras. 

\begin{flushright}
\LAP-A-507/95 \\
hep-th-9503042\\
February 1995
\end{flushright}

\end{titlepage}
\setcounter{footnote}{0}

\sect{Introduction}

In recent years, we have seen a great activity in the area of 
extended conformal algebras, i.e. algebras that contain the 
Virasoro algebra as a subalgebra, see e.g. \cite{BoSc}. Examples of this are  
the well known Kac-Moody algebras and superconformal algebras, 
but also the more complicated \cw\,algebras, introduced by Zamolodchikov 
in 1985, \cite{Za}. 
Apart from their inherent interest as mathematical objects, 
such algebras are of interest in many different fields of physics, 
such as the study of integrable hierarchies \cite{inthi}, 
string theory \cite{strth}, Toda theories \cite{toda}, 
quantum Hall effect \cite{Hall}, etc. 

Different methods have been developed for the construction of extended 
conformal algebras. One method is the direct construction, 
essentially the solution, 
using algebraic computation, of a set of consistency equations 
for a prescribed set of fields, see e.g. \cite{Ka}. Another method for 
constructing $W$-algebras is the coset method, the
generalization of the well-known coset construction in conformal field theory,
for a review see e.g. \cite{BoSc}. 
However, one of the most 
powerful methods of 
constructing extended conformal algebras is the hamiltonian reduction. The 
starting point here is an affine Lie (super)algebra, on which one imposes 
a suitably chosen set of constraints. The reduced algebra is then an 
extended conformal algebra 
\cite{BaTjVDr,BaFeFoRaWi,BeOo,FrRaSo}. The quantum version of this reduction 
has recently been done using BRST techniques 
\cite{KoSt,FeFr2,BoMcPi,FrKaWa,BoTj,SeThTr,MaRa}.

It has recently been shown \cite{DeFrRaSo} that in certain cases one can 
impose extra constraints on a \cw\ algebra obtained by hamiltonian reduction, 
and get yet 
another \cw\ algebra; this procedure is called secondary hamiltonian 
reduction. In this paper we show how to quantize this procedure, thus 
performing the secondary quantum hamiltonian reduction. 

As an application of the technique developed, we find the generalized 
quantum Miura transformation corresponding to the secondary reduction. 
This secondary quantum Miura transformation leads to 
a large number of new realizations of \cw\ algebras. 

As another application, we show that the secondary quantum hamiltonian 
reduction yields a general method of constructing linearizations of \cw\ 
algebras;
For a large class of \cw\ algebras, we find that we can use the secondary 
quantum 
hamiltonian reduction to construct linearizations in a systematic 
way. 
Recall that a linearization of a \cw\ algebra, introduced in \cite{BeKrSo}
is the embedding of that \cw\ algebra into a larger algebra which is 
equivalent to a linear algebra. 

This paper is organized in the following way: 
In section \ref{sec:pr}, we briefly remind the reader of the primary 
hamiltonian reduction; section \ref{sec:pcr} is a description of the classical 
reduction, while section \ref{sec:primared} gives a brief account of the 
quantum reduction. In section \ref{sec:scr} we briefly recall the classical 
secondary reduction.
These sections are included with the purpose of keeping the 
paper reasonably self-contained.

Sections \ref{sec:specseq} and \ref{sec:qhr} are the two main sections 
of the paper. In these sections we find the cohomology corresponding to the 
secondary quantum hamiltonian reduction. In section \ref{sec:specseq} we 
use the theory of spectral sequences to show that for triples \cg, 
$\ch'$ and $\ch$ with $\ch'\subset\ch$, and satisfying certain supplementary 
conditions, the secondary reduction of $\cw(\cg,\ch')$ gives as a result 
$\cw(\cg,\ch)$. In section \ref{sec:qhr} we give a method to find explicitly
the cohomology corresponding to $\cw(\cg,\ch)$, i.e. to find 
expressions for the generators of $\cw(\cg,\ch)$ in terms of the generators 
of $\cw(\cg,\ch')$. 
Furthermore we describe the generalized quantum Miura transformation 
corresponding to the
secondary quantum hamiltonian reduction, a transformation which gives us 
numerous new realizations of \cw\ algebras. Section \ref{exa:1} is an example
of the secondary quantum hamiltonian reduction. 

The main results of these sections are collected in theorem
\ref{the:coh} page \pageref{the:coh}, theorem \ref{the:expcoh} page 
\pageref{the:expcoh}, and theorem \ref{the:miura} page \pageref{the:miura}. 

In section \ref{sec:lin} we show how to use the technique, developed in the 
preceding sections, to 
linearize \cw\ algebras. Using the secondary quantum hamiltonian 
reduction we show that we can embed many \cw\ algebras into 
larger algebras, which are 
equivalent to linear algebras. 

Finally we have included two appendices, one on spectral sequences and one on 
the use of modified gradings in the hamiltonian reduction. 

For all explicit calculations, we have used the OPE mathematica package of 
K. Thielemans
\cite{KT}.

\sect{Hamiltonian Reductions: a Reminder}
\label{sec:pr}
\subsection{The Classical Case}
\label{sec:pcr}

\indent

Let us briefly recall the construction of classical \cw(\cg,\ch) algebras as 
they appear in the context of Hamiltonian reduction \cite{BaFeFoRaWi}. We 
start with a Lie algebra $\cg$ with generators $t^a$ and inner
product $g^{ab} = \langle t^a t^b\rangle$. Furthermore we consider a regular 
subalgebra $\ch\subset\cg$, and we denote the generators of the principal 
$s\ell(2)$ of \ch~by $\{ M_+,M_0,M_- \}$. $M_0$ defines a 
grading $gr(\cdot)$ of $\cg$
$$
\cg = \cg_- + \cg_0 + \cg_- = \sum_m \cg_m
$$ 
where $m$ is the eigenvalue under the operator $\ad(M_0)$. We denote the 
affine Lie algebra corresponding to \cg~as $\cg^{(1)}$, and we write the 
affine current $J$ in the form
$$
J(z) = J^a(z)\ t_a. 
$$
$t_a = g_{ab}t^b$ is an element of the dual algebra, $g_{ab}$ is defined by 
$g_{ab}g^{bc} = \D_a^c$. We will use greek letters as indices for currents 
with negative grades, and barred greek letters for currents with non-negative
grades, i.e. $J^\A:\,t_\A\in\cg_-$ and 
$J^{\bar{\A}}:\,t_{\bar{\A}}\in\cg_0\cup\cg_+$.
The constraints that we want to impose are 
\ben
\label{eq:y}
(J^\A(z) - \chi^\A) = 0 \quad (t^\A \in \cg_-)
\een
$\chi^\A=\chi(J^\A)$ define the set of constraints and 
$M_-=\chi^\A t_\A$.
These constraints are chosen such that they are first class. 
This means that the Poisson bracket of two 
constraints is weakly zero (i.e. one finds zero when imposing the constraints 
after 
computation of  the Poisson bracket). We can then apply the general technique 
developed by Dirac to take care of these constraints: 

The first class constraints generate gauge transformations, i.e. they are 
associated to a group of symmetries of the physical states of the theory. To 
eliminate 
this symmetry, one imposes new constraints (gauge fixing constraints) in 
such a way that the set of all constraints becomes 
second class (i.e. it is no more first class), and the matrix formed by the 
Poisson brackets 
of two constraints $C_{ij}(t_1,t_2)=\{\Phi_i(t_1),\Phi_j(t_2)\}$ is 
invertible. As these constraints must not interfere with 
the physical contents of the theory, one constructs new brackets (called Dirac 
brackets) 
with the help of the Poisson brackets and the inverse $C^{ij}$ of $C_{ij}$:
\ben
\{X(z),Y(w)\}_D= \{X(z),Y(w)\}- 
\int dt_1dt_2\ \{X(z),\Phi_i(t_1)\} C^{ij}(t_1,t_2) \{\Phi_j(t_2),Y(w)\}
\een
where $\Phi_i(t)$ are the (second class) constraints.
These Dirac 
brackets are defined such that any quantity has (strongly) zero Dirac 
brackets with any of the 
constraints. In other words, we have decoupled the constraints from the other 
physical 
quantities. 

In the case of the constraints (\ref{eq:y}), it can be shown that 
one can choose gauge-fixing constraints 
such that the remaining generators correspond to the highest 
weight components with respect to the embedded $s\ell(2)$: The gauge-fixed 
current $J_{gf}$ is of the form 
$J_{gf} = \chi^\A t_\A + J^{\bar{\imath}}t_{\bar{\imath}}$ 
with $[M_+, t_{\bar{\imath}}]=0$.
As the constraints are decoupled from the other physical quantities,
it is clear that the Dirac bracket of two $J^{\bar{\imath}}$'s will 
close (polynomially) on the $J^{\bar{\imath}}$'s. These Dirac brackets 
realize the \cw\ algebra \cw(\cg,\ch).

To get a realization of the $W$ generators as polynomials of the currents 
$J^a$, one uses the gauge transformations generated by the first class 
constraints 
\ben
\begin{array}{ll}
[J^a(w)]^g= & J^a(w) +\int dz\ \eps_\A(z)\{J^\A(z), J^a(w)\} \\
& +\frac{1}{2!} \int dzdx\ \eps_\A(z)\eps_\B(x)
\{J^\B(x),\{J^\A(z), J^a(w)\}\} +\ldots
\end{array}
\een
to fix $[J(w)]^g$ to be of the form
\ben
[J(z)]^g= \chi^\A t_\A + W^{\bar{\imath}}(z) t_{\bar{\imath}}
\een
where the $W^{\bar{\imath}}(z)$ are polynomials in the $J^a$'s and realize 
also the \cw\ algebra when using the {\it Poisson} brackets. 

It can also be shown that one can realize the \cw\ algebra by using the zero 
grade generators only: starting with 
$J(z)= \chi^\A t_\A + J^{\bar{\A}_0}(z)t_{\bar{\A}_0}$ where 
$[M_0, t_{\bar{\A}_0}]=0$ and doing the gauge transformations as above, one 
gets the $W^{\bar{\imath}}(z)$ as polynomials in the $J^{\bar{\A}_0}$'s. This 
transformation is called the (classical) Miura transformation.

Finally, let us note that the choice (\ref{eq:y}) of constraints is not 
unique. We
introduce a new grading operator $H = M_0 + U$, where $U$ is an element of the 
Cartan subalgebra. $H$ defines a grading which we write as 
$$
\cg = \cg_-' + \cg_0' + \cg_+' = \sum_n \cg_n'.
$$
In \cite{U1,FrRaSo} it was shown that the constraints obtained by replacing 
$\cg_-$ by $\cg_-'$ in equation (\ref{eq:y}) leads to the same classical 
algebra, 
$\cw(\cg,\ch)$, 
if $U$ commutes with the $s\ell(2)$ algebra and ''respects" the highest 
weights, i.e. satisfies the non-degeneracy condition
$$
\mbox{ker }ad(M_+)\cap \cg'_-={0}
$$

\subsection{Primary Quantum Hamiltonian Reduction}
\label{sec:primared}

This section is intended as a brief recapitulation of the method developed 
in \cite{BoTj} of quantum hamiltonian reduction. 
We want to quantize the Hamiltonian reduction which we have presented in the 
previous section. 
To do this, we will use the BRST formalism, which is a standard
procedure in the framework of constrained systems, see e.g. \cite{KoSt}.

For each constraint we introduce a ghost-antighost pair $(c_\A,b^\A)$. 
Corresponding to these constraints we define a BRST operator $s$: 
\bea
\label{eq:s}
s(\phi)(w) & = & \oint_w\ dz\ j(z) \phi(w) \nn
j(z) & = & (J^\A(z) - \chi^\A)c_\A(z) 
+ \hf {f^{\A\B}}_\G b^\G(z) c_\B(z) c_\A(z)
\eea
The quantized \cw~algebra $\cw(\cg,\ch)$ is then 
$$
\cw(\cg,\ch) = H^0(\OM;s)
$$
where $\OM$ is the operator product algebra generated by the affine currents, 
the ghosts and anti-ghosts, and their derivatives and normal ordered products.
As usual when using BRST quantization, one defines a set of modified 
generators
$$
\hat{J}^\A(z) = s(b^\A)(z) +\chi^\A= J^\A(z) + {f^{\A\B}}_\G b^\G(z) c_\B(z)
$$
and it turns out that it is useful to modify in a similar way the 
non-constrained generators: 
$$
\hat{J}^{\bar{\A}}(z) = J^{\bar{\A}}(z) + {f^{\bar{\A}\B}}_\G (b^\G c_\B)_0(z)
$$
where we have introduced the normal ordered product of fields 
$A_j(z)$ ($j=1,2,...)$:
\ben
(A_1A_2)_0(w)= \oint_w \frac{dz}{z-w} A_1(z)A_2(w)
\mb{and} (A_n\dots A_1)_0(w) =(A_n(A_{n-1}\dots A_1)_0)_0(w)
\een
With these definitions the space $\OM^\A$ generated by $\hat{J}^\A$ and $b^\A$ 
is actually a subcomplex (i.e. $s(\OM^\A) \subset \OM^\A$) with trivial 
cohomology 
$H^n(\OM^\A;s) = \D_{n,0}\,\C$. The space $\OM_{red}$ generated by the 
``hatted'' non-constrained generators $\hat{J}^{\bar{\A}}$ and the ghosts 
$c_\A$ is also a subcomplex, and in fact using the version of the 
K\"unneth formula which 
was shown in \cite{BoTj}, one can show that we need not 
consider the trivial subcomplexes $\OM^\A$ since
\bea 
\label{eq:kunneth}
H^*(\OM;s) & \cong & 
H^*\left ( \OM_{red}\otimes (\otimes_\A \OM^\A ) ; s \right ) \, \cong \, 
H^*(\OM_{red};s) \otimes ( \otimes_{\A} H^*(\OM^\A;s) ) \nn
& \cong & H^*(\OM_{red};s)
\eea

In order to actually calculate this cohomology and find explicit expressions 
for the generators of $\cw(\cg,\ch)$, one splits the BRST operator $s$ into 
two nilpotent anticommuting operators $s_0$ and $s_1$ defined by the currents 
\bea 
j_0(z) & = & -\chi^\A c_\A(z) \nn
j_1(z) & = & J^\A(z) c_\A(z) + \hf {f^{\A\B}}_\G b^\G(z) c_\B(z) c_\A(z)
\eea
Corresponding to these two operators, we can define a bigrading of the complex 
$\OM$ as a combination of the ghostnumber and the grading $gr(\cdot)$: 
\bea 
J^a & : & (m,-m) \nn
b^\A & : & (m,-m-1) \label{quoi}\\
c_\A & : & (-m,m+1) \nonumber
\ena 
where $m$ is the grade of $t_a$ respectively $t_\A$. With this definition 
$s_0$ has bigrade $(1,0)$ and $s_1$ has bigrade $(0,1)$. 

Using the technique of spectral sequences (see section \ref{sec:specseq} and 
appendix \ref{ap:a}), one can show that there is a vector space isomorphism 
$$ 
H^n(\OM_{red};s) \cong H^n(\OM_{red};s_0) \cong \D_{n,0}\,\OM_{hw},
$$
where $\OM_{hw}$ is generated by the hatted generators that are highest
weight under the embedded $s\ell(2)$, i.e. 
$\hat{J}^{\bar{\A}} \in \OM_{hw}$ iff $[M_+,t_{\bar{\A}}]=0$.

In order to actually find the elements in $H^0(\OM_{red};s)$, i.e. the 
generators of $\cw(\cg,\ch)$, one can use the tic-tac-toe construction with 
$\OM_{hw}$ as starting point. For every highest weight generator 
$\hat{J}^{\bar{\A}}_{hw}$ we can construct the generator $W^{\bar{\A}}(z)$: 
$$
W^{\bar{\A}}(z) = \sum_{\ell=0}^{m} W^{\bar{\A}}_\ell (z)
$$
where $m$ is the grade of $\hat{J}^{\bar{\A}}_{hw}$, 
$W^{\bar{\A}}_0(z) = \hat{J}^{\bar{\A}}_{hw}(z)$, and 
$s_1(W^{\bar{\A}}_{\ell}) + s_0(W^{\bar{\A}}_{\ell+1}) = 0$.
We find that the bi-grade of $W^{\bar{\A}}_\ell$ is 
$(m-\ell,\ell-m)$, and since $s_1$ vanishes on terms with bigrade $(0,0)$
we see that the sequence stops at $\ell=m$. 
It is easy to verify that $s(W^{\bar{\A}})=0$. 

In principle the operator product algebra of the $W$'s close only modulo 
$s$-exact terms. 
However, there are no elements in $\OM_{red}$ with negative ghostnumber, 
thus there are no $s$-exact terms with zero ghostnumber, and therefore 
the operator products of the $W$'s close exactly. 

The operator product expansion (ope) preserves the grading, which implies that 
the operator product expansions of the zero grade part of the generators 
must give the same algebra as the ope's of the full generators, i.e. 
the map 
$W^{\bar{\A}} = \sum_{\ell=0}^{m} W^{\bar{\A}}_\ell 
\rightarrow W^{\bar{\A}}_m$ must be an algebra isomorphism\footnote{ 
actually this argument shows only that the map is an algebra homomorphism, 
but one can prove that the map is also injective.}. This defines a realization 
of $\cw(\cg,\ch)$ in terms of the algebra $\hat{\cg}^{(1)}_0$, the 
algebra generated by the ``hatted'' grade zero affine currents. 
This is known as the 
quantum Miura transformation, and can be used to define a free field 
realization 
of $\cw(\cg,\ch)$ by using the Wakimoto construction \cite{FeFr} 
to write the generators 
of $\hat{\cg}^{(1)}_0$ in terms of free fields. 

Just as in the classical hamiltonian reduction, we can modify the grading 
operator $M_0$ by adding a $U(1)$ current obeying the non-degeneracy 
condition. In that case, the modified grading operator 
$H=M_0+U$ will lead to a modification of the BRST operator (\ref{eq:s}). 
One finds, that the calculation of the cohomology leads to an 
equivalent but different ("twisted") realization of $\cw(\cg,\ch)$. 

\subsection{Classical Secondary Reductions}
\label{sec:scr}

\indent

First, we briefly recall the framework of secondary reductions as they appear 
in the 
classical case. We start with a $\cw(\cg,\ch')$ algebra (defined as in section 
1), with $\ch'$ 
a regular subalgebra of $\cg$. We suppose now that there is another regular 
subalgebra 
$\ch$ such that $\ch'\subset\ch$. Since $\ch'$ is embedded in $\ch$, it is 
natural to 
wonder whether the $\cw(\cg,\ch')$ algebra can be related to $\cw(\cg,\ch)$. 
In fact, 
considering the constraints associated to both $\cw$-algebras, it is clear 
that we have to 
impose more constraints on $\cw(\cg,\ch')$ to get $\cw(\cg,\ch)$; for 
instance, the 
number of primary fields (which is directly related to the number of 
constraints) is lower 
in $\cw(\cg,\ch)$ than in $\cw(\cg,\ch')$. These (further) constraints will 
be imposed on 
$W$ fields themselves, so that we will gauge a part of the $\cw(\cg,\ch')$ 
algebra. In 
\cite{DeFrRaSo}, it has been proved:

\begin{theo}
Let $\cg=s\ell(N)$ and let $\ch'$ and $\ch$ be two regular 
subalgebras of $\cg$ 
such that 
\be
\ch'\subset\ch
\ee
Then, there is a set of constraints on the $\cw(\cg,\ch')$ algebra such that 
the (associated) 
Hamiltonian reduction of this algebra leads to the $\cw(\cg,\ch)$ algebra. 
We will represent this secondary reduction as 
\be
\cw(\cg,\ch')\ \rightarrow\ \cw(\cg,\ch)
\ee
\end{theo}
The proof of this theorem relies on a general property of the Dirac brackets, 
which can be stated as follows: 

We start with a Hamiltonian theory on which we impose constraints.  
Instead of considering directly the 
complete set of second class constraints, we can divide this set into 
several subsets (of 
second class constraints) and compute the Dirac brackets at each steps 
(using the Dirac 
bracket of the previous steps as initial Poisson brackets). Then the last 
Dirac 
brackets do not 
depend on the partition of the second class constraints set we have used. 

Thus, coming back to our $\cw$-algebras, it is sufficient to find a gauge 
fixing for the 
$\cw(\cg,\ch')$ algebra such that the corresponding set of second class 
constraints is embedded into the 
set of second class constraints for $\cw(\cg,\ch)$ as soon as 
$\ch'\subset\ch$. Indeed, 
with such an embedding, it is clear that the constraints one will impose on 
the 
$\cw(\cg,\ch')$ generators will just be the constraints related to $\ch$ that 
are not in the 
subset associated to $\ch'$. Such a gauge has been explicitly constructed 
in \cite{DeFrRaSo} for  $\cg=s\ell(N)$. 
Because of the generality of the property of Dirac brackets, and considering 
the 
construction of orthogonal and symplectic algebras from the (folding of) 
unitary ones \cite{FrRaSo2}, it is clear that the theorem is also true 
for the other 
classical Lie algebras. 

We present below a quantization of the secondary reduction using the BRST 
formalism. 
Note that the BRST operator involves only first class constraints, so that 
the quantization is 
not straightforward: in the classical case, we have to embed the sets of 
second class 
constraints one into the other, while in the quantized version it is the 
sets of first class 
constraints that we will embed.

\sect{Quantum Secondary Reductions: \protect \\ Algebra Isomorphism}
\label{sec:specseq}

In order to show that $\cw(\cg,\ch)$ can be obtained from a secondary 
hamiltonian reduction of $\cw(\cg,\ch')$, we will use the theory of spectral
sequences. For a good introduction
see e.g. \cite{Mc}; in appendix \ref{ap:a} we give a brief 
description of some main points in the theory. 

Assume that we have a Lie algebra \cg\ and two regular subalgebras 
$\ch'$ and \ch\ with $\ch' \subset \ch$, as in the previous section.
The principal $s\ell(2)$ subalgebra of $\ch'$ is denoted by 
$\{ M'_-,M'_0,M'_+\}$, while the principal $s\ell(2)$ subalgebra of \ch~ is 
$\{ M_-,M_0,M_+\}$. 
The eigenvalues of the operator $\ad(M_0)$ defines a natural grading of \cg:
\ben 
\cg \, = \, \cg_- + \cg_0 + \cg_+ \, = \, \sum_{m} \cg_m. 
\een 
A second grading is defined by $M_0'$, but as described in appendix
\ref{sec:u1} we can use the more general, but equivalent, grading operator 
$\ad(H') = \ad(M_0' + U)$. 
We write the corresponding grading as: 
\ben
\cg \, = \, \cg'_- + \cg'_0 + \cg'_+ \, = \, \sum_{n} \cg'_{n}.
\een

We assume the gradings to be integer, and call $H$ the corresponding grading 
operator\footnote{
For the algebras which we consider, it is always possible to choose an integer
grading.}. We wish to constrain the negative grade parts of the two 
algebras, $\cg'_-$ and $\cg_-$ respectively. 
To each generator $t^\A\in\cg_-$ corresponds one of the first class 
constraints of the type $\phi^\A(z)=J^\A(z)-\chi^\A=0$ where the $\chi^\A$ are 
constants. We denote by $\Phi$ the set of these first class constraints, and 
similarly $\Phi'$ denotes the set of constraints corresponding to $\cg_-'$. 
We assume that we can choose the constraints in such a way that 
$\Phi' \subset \Phi$, and we note that this implies that the following two
conditions are satisfied: 
\begin{itemize}
\item[1)] the set of simple roots of $\ch'$ can be 
chosen to be a subset of the set of simple roots of \ch, and 
\item[2)] $\cg_-' \subset \cg_-$. 
\end{itemize}
The classification of triples \cg, $\ch'$, and \ch\ satisfying these 
conditions are given in appendix \ref{sec:u1}. Note that although the ``usual''
constraints (defined by the grading $\ad(M_0)$) obeying $\Phi'\subset\Phi$ 
are very few, the use of modified gradings as described in the appendix 
gives us a large class of triples that satisfy $\Phi'\subset\Phi$.
 
Let us introduce the notation for the indices: 
\bea
\label{eq:indices}
t_A & \in & \cg_- \quad \quad \, t_{\bar{A}} \in \cg_0 \cup \cg_+ \nn
t_a & \in & \cg_-' \quad \quad t_{\bar{a}} \in \cg'_0 \cup \cg'_+ \nn
t_\A & \in & \cg_- \setminus \cg_-' 
\eea

Note that the generators $t_\A$ must have grade zero with respect to 
$H'$: $t_\A\in\cg'_0$ (if $t_\A\in\cg'_{+}$ then we can find a 
$t_{\bar{\A}}\in\cg'_{-}$, corresponding to a generator in 
$\cg_-'\setminus\cg_-$; but this is in contradiction with the condition 2) 
above). 
Note also that $t_\A$ is a highest weight generator under the embedding 
of $\{ M'_-,M'_0,M'_+\}$. To show this, assume a root basis, and let 
$\A_1,\A_2,\ldots,\A_n$ denote the 
simple roots of $\ch'$, and let $\A_1,\A_2,\ldots,\A_n,\B_1,\B_2,\ldots,\B_m$ 
denote the simple roots of \ch. We can write $M'_+ = \sum a^i t_{\A_i}$. 
On the other hand, since $t_\A$ has grade zero under $H'$ we find that 
we can write $\A = - \sum n_j \B_j$. This shows that $[M'_+,t_\A]=0$. 
 
We can write the constraints in the form: 
\bea
\label{eq:con'}
\phi^a(z) & = &  \quad J^a(z) - \chi^a = 0, \quad\quad \phi^a \in \Phi' \\
\label{eq:con}
\phi^A(z)  & = & \quad J^A(z) - \chi^A = 0, \quad\quad \phi^A \in \Phi
\eea
where $M'_- = \chi^a t_a$ and $M_- = \chi^A t_A$. 

Corresponding to the constraints (\ref{eq:con'}) we
introduce ghosts $c_a$ and anti-ghosts $b^a$, and we define the BRST current 
$j'$ by (see eq. (\ref{eq:s}))
$$
j'(z) = (J^a(z) - \chi^a) c_a(z) + \hf {f^{ab}}_c b^c(z) c_b(z) c_a(z). 
$$
Similarly we introduce ghosts $c_A$ and anti-ghosts $b^A$ corresponding to 
(\ref{eq:con}). The set of ghosts $c_a$ is a subset of the set $c_A$ and 
$b^a$ is a subset of $b^A$; in fact the set  
$c_A$ is the union of the set $c_a$ and the set $c_\A$, and the set $b^A$ is 
the union of the set $b^a$ and the set $b^\A$. 
The BRST current $j$ is defined by 
$$
j(z) = (J^A(z) - \chi^A) c_A(z) + \hf {f^{AB}}_C b^C(z) c_B(z) c_A(z). 
$$
We define the current $j''$ by $j = j' + j''$, and we find 
$$
j''(z) = j(z) - j'(z) = (J^\A(z) - \chi^\A) c_\A(z) + 
\hf ({f^{AB}}_C b^C(z) c_B(z) c_A(z) - {f^{ab}}_c b^c(z) c_b(z) c_a(z)) 
$$

Corresponding to the currents $j,\,j'$, and $j''$ we define operators $s,\,s'$,
and $s''$ by 
$$
s \phi(w) = \oint_w dz \, j(z) \phi(w), \nn
$$
and similarly for $s'$ and $s''$. 

Using $t_\A \in \cg'_0$ and $t_a \in \cg'_-$ we can show that terms 
of the form ${f^{ab}}_\G b^\G c_b c_a$, ${f^{a\B}}_\G b^\G c_\B c_a$, and 
${f^{\A\B}}_c b^c c_\B c_\A$ vanish. This means that we can write: 
\bea
\label{eq:j''} 
j''(z) & = & (J^\A(z) - \chi^\A) c_\A(z) \nn 
& & +
\hf ({f^{\A\B}}_\G b^\G(z) c_\B(z) c_\A(z) + 
{f^{a\B}}_c b^c(z) c_\B(z) c_a(z) + 
{f^{\A b}}_c b^c(z) c_b(z) c_\A(z) ) \nn
& = & (\tilde{J}^\A(z) - \chi^\A) c_\A(z) + 
\hf {f^{\A\B}}_\G b^\G(z) c_\B(z) c_\A(z)
\eea
where in the last line $\tilde{J}^\A$ is defined by 
$\tilde{J}^\A(z) = s'(b^\A)(z) + \chi^\A 
= J^\A(z) + {f^{\A b}}_c b^c(z) c_b(z)$. 
Note that since 
$J^\A$ is a highest weight generator with grade zero under the grading 
operator $H'$, $\tilde{J}^\A$ is actually a generator 
$W^\A$ of the algebra $\cw(\cg,\ch')$, so we can alternatively write 
\ben 
\label{eq:j2}
j''(z)  = (W^\A(z) - \chi^\A) c_\A(z) + 
\hf {f^{\A\B}}_\G b^\G(z) c_\B(z) c_\A(z)
\een

Let \ca\ denote the algebra generated by the currents as well as their
derivatives and normal ordered products. $\LM'$ is the algebra generated by the
ghosts $c_a$ and the anti-ghosts $b^a$ (and their derivatives and normal
ordered products), $\LM''$ is generated by $c_\A$ and $b^\A$, and $\LM$ is
generated by $c_A$ and $b^A$. Note that $\LM = \LM' \otimes \LM''$. 
The algebras $\ca\otimes \LM'$ and $\ca\otimes \LM$ are graded by 
ghost numbers, and we know \cite{KoSt,FeFr2,BoTj} that 
\bea 
\cw(\cg,\ch) & \cong & H^0(\ca\otimes \LM ; s),     \nn
\cw(\cg,\ch') & \cong & H^0(\ca\otimes \LM' ; s').    
\eea

We can define a bigrading on the algebra 
$\OM=\ca\otimes\LM:\, \OM = \sum_{p,q} \OM^{p,q}$ 
such that 
$s'$ has bigrading $(1,0)$ and $s''$ has bigrading $(0,1)$, namely: 
\bea
J    & : & (0,0)   \nn
c_a  & : & (1,0) \nn
c_\A & : & (0,1) \\
b^a  & : & (-1,0) \nn
b^\A  & : & (0,-1) \nonumber
\eea
Be careful that this bigrading is not the bigrading used in the previous 
section: it is based on two
ghostnumbers, while for primary reductions, the bigrading is based on the 
gradation of \cg, and on
one ghostnumber (see eq. (\ref{quoi})).
Define 
\bea
\hat{J}^A \,=\, s(b^A)+\chi^A & = & J^A + {f^{AB}}_C b^C c_B, \nn
\hat{J}^{\bar{A}} & = & J^{\bar{A}} + {f^{\bar{A}B}}_C b^C c_B, \nonumber
\eea 
$J^A_{ghost} \equiv {f^{AB}}_C b^C c_B$ is the ghost realization of the
constrained part of the algebra. 
We will use as basis of $\OM$ the set of ``hatted" currents and the 
ghosts and anti-ghosts $\{ \hat{J}, c, b \}$. For each index 
$A$ the algebra $\OM^A$ generated by $\hat{J}^A$ and $b^A$ is an 
$s$-subcomplex 
with trivial cohomology \cite{BoTj}
$$
H^n( \OM^A;s ) \cong \D_{n,0}\, \C. 
$$
Define $\bar{\OM}_{red}$ to be the algebra generated by 
$\{ \hat{J}^{\bar{A}}, c_A \}$. As in eq. (\ref{eq:kunneth}) we find that 
\bea 
H^*( \OM ; s) 
& \cong & H^* \left ( \bar{\OM}_{red} 
\otimes \left ( \bigotimes_{A} \OM^A \right ) ; s \right )  \quad \cong \quad 
H^*( \bar{\OM}_{red} ;s ) \otimes \left ( \bigotimes_{A} H^*(\OM^A ; s)  
\right ) \nn
& \cong & H^*( \bar{\OM}_{red} ;s ) \otimes \left ( \bigotimes_{A} \C \right ) 
\quad \cong \quad H^*( \bar{\OM}_{red} ;s ) 
\eea
i.e. we can reduce the problem to finding the cohomology of $\bar{\OM}_{red}$, 
ignoring the trivial subcomplexes $\OM^A$. In our case it will turn out 
to be convenient to perform this reduction only partly, in the sense that 
we will use the K\"unneth formula only to extract the subcomplexes $\OM^a$. 
We will therefore define $\OM_{red}$ to be the subcomplex generated by 
$\{ \hat{J}^{\bar{a}}, c_a \}$ and $\LM''$. The full complex $\OM$ can be 
written in the form 
$$
\OM \cong \OM_{red} \otimes \left ( \bigotimes_a \OM^a \right ) 
$$
and we use the K\"unneth formula to find 
\ben 
\label{eq:red}
H^n( \OM ; s) \cong H^n( \OM_{red} ;s ) 
\een 

Now we make a change of basis. As new basis we choose the 
currents $\tilde{J}^{\bar{a}} = J^{\bar{a}} + {f^{{\bar{a}} b}}_c b^c c_b$, 
the ghosts $c_a$ and $c_\A$, and and anti-ghosts $b^\A$. We denote 
the space generated by $\tilde{J}^{\bar{a}}$ and $c_a$ by $\Gamma$, 
so we have $\OM_{red} =\Gamma\otimes\LM''$. 
Note that the cohomology of
$\Gamma$ with respect to the operator $s'$ is the
$\cw(\cg,\ch')$ algebra: 
$$
H^n(\Gamma ; s' ) \cong \D_{n,0} \, \cw(\cg,\ch'). 
$$
Note also that we have 
$(\Gamma \otimes \LM'')^{p,q} = \Gamma^{p,0} \otimes (\LM'')^{0,q}$. 

We can now consider the spectral sequence corresponding to the double complex 
$(\OM_{red} ; s'; s'')$. The spectral sequence is a sequence of complexes 
$(E^{p,q}_r ; s_r)$, such 
that 
\bea 
E^{p,q}_0 & = & (\OM_{red})^{p,q} \nn 
E^{p,q}_{r+1} & = & H^{p,q}( E_r ; s_r ) \, = \, 
\frac{ E^{p,q}_r \cap \ke(s_r) }{ E^{p,q}_r \cap \im(s_r) }  
\eea
where $s_r$ is a nilpotent operator of bigrade $(1-r,r)$, $s_0 = s'$ and 
$s_1 = [s'']$. The operators $s_r$ for $r\geq 2$ are defined in appendix 
\ref{ap:a}.
The notation $s_1 = [s'']$, is to be interpreted as 
$s_1 ([x]) = [s''(x)]$ of a given $[x] \in E_1$. 
This is well-defined because 
$$
[s''(x+s'(y))] = [s''(x) + s''(s'(y))] = [s''(x) - s'(s''(y))] = [s''(x)].
$$ 
It is now possible to show that if the spectral sequence collapses, i.e. if 
there exists $R$ such that $E_r = E_R$ for $r \geq R$, then\footnote{
Actually this condition is sufficient but not necessary; it can in fact 
be relaxed considerably, see appendix \ref{ap:a}.} 
we have 
\ben
\label{eq:th1}
E_\infty^{p,q} \cong F^q H^{p+q} / F^{q+1} H^{p+q}
\een
where $E_\infty = E_R$ and $F^q H$ is a filtration on the cohomology 
$H(\OM_{red};s)$ defined by 
\ben
F^q H^{p+q} \, = \, H^{p+q} (\bigoplus_{i\geq 0} (\OM_{red})^{p-i,q+i} ;s)
\, = \, \frac{(\bigoplus_{i\geq 0} (\OM_{red})^{p-i,q+i}) \cap \ke s}
{(\bigoplus_{i\geq 0} (\OM_{red})^{p-i,q+i}) \cap \im s} \label{toto}
\een
thus we can in principle reconstruct the cohomology $H(\OM_{red};s)$ from the 
spectral sequence, on the condition that we can reconstruct $H(\OM_{red};s)$ 
from the quotient spaces $F^q H^{p+q} / F^{q+1} H^{p+q}$. 

The first element in the spectral sequence is  
$$
E_0^{p,q} = (\OM_{red})^{p,q} = (\Gamma\otimes\LM'')^{p,q}.
$$
For the second element we find 
\ben
\label{eq:coho}
E_1^{p,q} \, = \, H^{p,q} (E_0; s_0) 
\, \cong \, H^p(\Gamma ; s' ) \otimes (\LM'')^{0,q}
\, \cong \, \D_{p,0} \, \cw(\cg,\ch')  \otimes (\LM'')^{0,q}.
\een
The third element in the spectral sequence is $E_2 = H(E_1,s_1)$. 
we find 
\ben 
\label{eq:e2}
E_2^{p,q} = H^{p,q}(E_1;s_1) 
= \D_{p,0} H^{p,q}(\cw(\cg,\ch') \otimes \LM''; [s''] )
\een

The spectral sequence collapses here, i.e. $[s'']$ is the last 
non-trivial operator in the sequence. In fact $E_2^{p,q}$ is nontrivial only
for $p=0$, and since $s_r$ has bigrade $(1-r,r)$ it is clear that $s_r$ is 
trivial for $r\geq 2$. 
We conclude that $E_r = E_2$ for any $r \geq 2$, and so 
$E_\infty = E_2$. Note that from equation (\ref{eq:j2}) it follows that if 
we restrict $s''$ to $\cw(\cg,\ch') \otimes \LM''$ then it maps into 
 $\cw(\cg,\ch') \otimes \LM''$, which means that we can replace 
$[s'']$ by $s''$ in equation (\ref{eq:e2}).   

From equation (\ref{eq:th1}) it follows that we have 
$$
F^q H^{p+q} / F^{q+1} H^{p+q} \cong E_2^{p,q} 
\cong \D_{p,0} H^{p,q}(\cw(\cg,\ch') \otimes \LM''; s'' ),
$$
where $F^q H^{p+q}$ is defined in equation (\ref{toto}).

$(\OM_{red})^{p,q}$ is trivial for $p < 0$, and therefore 
$\bigoplus_{i>0} (\OM_{red})^{p-i,q+i}$ is trivial for $p<0$. 
Using lemma \ref{eq:recH} of appendix \ref{ap:a}, one can verify 
that this implies that 
$H(\ca\otimes\LM ; s) \cong E_2$, and that this isomorphism is in fact an 
algebra isomorphism. 

Let us collect the results of this section in the following 
\begin{theo}
\label{the:coh}
Given two \cw~algebras $\cw' = \cw(\cg,\ch')$ and $\cw = \cw(\cg,\ch)$ with 
$\ch'\subset\ch$. If we can find sets of first class constraints 
$\Phi'$ and $\Phi$ (where $\cw'$ is the result of imposing the set of 
constraints $\Phi'$ on $\cg^{(1)}$ and \cw~is the result of imposing $\Phi$ 
on $\cg^{(1)}$)
such that $\Phi'\subset\Phi$, then: 
\begin{itemize}
\item[1)] It is possible to perform a secondary quantum hamiltonian 
reduction on $\cw'$. This secondary reduction consists of imposing 
a set $\Phi''$ of first class constraints on $\cw'$. There is a simple 
one-to-one correspondence between the constraints $\Phi''$ imposed on $\cw'$, 
and the ``missing'' constraints $\Phi\setminus\Phi'$. 
\item[2)] Let \ca\, be the algebra generated by currents in $\cg^{(1)}$ and 
their derivatives and normal ordered products, and let $\LM'$ be the algebra 
generated by the ghosts and anti-ghosts corresponding to the constraints 
$\Phi'$. If we denote by $s'$ and $s$ the BRST operators corresponding to 
the quantum hamiltonian reduction leading to $\cw'$ and \cw~respectively, then
the BRST operator that corresponds to the secondary quantum hamiltonian 
reduction of $\cw'$ is $[s-s']\equiv [s'']$. Considering $\cw'$ as the 
cohomology $H^0(\ca\otimes\Lambda';s')$, $[s'']$ on an element 
$[x] \in \cw'$ is defined by $[s'']([x]) \equiv [s''(x)]$. 
\item[3)] Let $\LM$ and $\LM''$ be the algebras generated by the ghosts 
corresponding to $\Phi$ and $\Phi''$ respectively. 
The result of the secondary hamiltonian reduction of $\cw'$ is 
$$
H^0 ( \cw(\cg,\ch') \otimes \LM'' ; [s''] ) \, \cong \,
H^0 ( H^0 ( \Gamma \otimes \LM'' ; s' ) ; s'') \, \cong \, 
H^0 ( \ca\otimes\LM ; s) \, \cong \, \cw(\cg,\ch).
$$
\end{itemize}
\end{theo}

\sect{Quantum Secondary Reduction: \hbox{Direct Calculation} }
\label{sec:qhr}

In section \ref{sec:primared} we explained briefly the primary 
quantum hamiltonian reduction of the affine Lie algebra $\cg^{(1)}$ that 
results in the \cw~algebra $\cw(\cg,\ch)$. Let us recall some of the main 
points of the procedure:
\begin{itemize}
\item[1)] The cohomology 
$H^0(\ca\otimes\LM;s)\equiv H^0(\OM;s)$ is isomorphic to the 
cohomology $H^0(\OM_{red};s)$, where 
$\OM_{red}$ is the space generated by the ``hatted'' 
unconstrained generators $\hat{J}^{\bar{\A}}$ and the ghosts $c_\A$. 
\item[2)] There is a vector space isomorphism between the space $\OM_{hw}$, 
generated by the hatted highest weight generators, and $\cw(\cg,\ch)$. 
\item[3)] We can use the tic-tac-toe construction to construct explicit 
realizations of the generators of \cw~in terms of the hatted unconstrained 
generators. The starting points for the tic-tac-toe construction are the 
elements of $\OM_{hw}$. 
\item[4)] We can show that there exists an algebra isomorphism between the 
algebra $\cw(\cg,\ch)$, and the algebra generated by the zero-grade part of the
$W$-generators as constructed by the tic-tac-toe method. 
This is the generalized quantum Miura 
transformation. 
\end{itemize}

In this section we will take the corresponding steps for the secondary quantum 
hamiltonian reduction. Among the consequences will be the secondary
quantum Miura transformation and a systematic method of linearization of 
\cw~algebras. 

Note that we have already found the BRST cohomology $H^0(\cw'\otimes\LM'';s'')$
to be identical to the algebra $\cw(\cg,\ch)$; the aim of this section is to 
construct concrete realizations of $\cw(\cg,\ch)$-generators from the 
generators of $\cw(\cg,\ch')$.

\subsection{Isomorphism between $\cw(\cg,\ch)$ and $H_0(\Gamma_{red},s)$}

The reduction of the $\cw'$ algebra is defined in terms of the grading 
$(H-H')$. The fact that this is actually a well-defined grading of the algebra 
follows from the fact that the simple roots of $\ch'$ has grade 1 both under 
$H$ and $H'$, which implies that $[M'_+,(H-H')]=0$ and therefore 
$\tilde{H}-\tilde{H'}$ is a generator of $\cw'$. The generators
to be constrained are $W^\A = \tilde{J}^\A$, which are just the generators 
with negative grade. 

Define $\Gamma=\cw(\cg,\ch')\otimes\LM''$.
Just as in the case of the primary reduction, we define ``hatted'' constrained 
generators by 
$$
\hat{W}^\A(z) = s''(b^\A)(z)+\chi^\A
$$
and we find that in fact $\hat{W}^\A = \hat{J}^\A$. 
For each $\A$, define 
$\Gamma^\A$ to be space generated by $\hat{W}^\A$ and $b^\A$. We note that 
$\Gamma^\A$ is a subcomplex with trivial cohomology: 
$$
H^n(\Gamma^\A ; s'') = \D_{n,0}\,\C
$$

In the primary quantum hamiltonian reduction, we saw in section 
\ref{sec:primared} that it was possible to split the complex $\Omega$ into 
a product of subcomplexes
$$
\Omega = \Omega_{red} \otimes \left ( \otimes_\A \Omega^\A \right ), 
$$
where the subcomplex $\Omega_{red}$ was generated by the ``hatted'' 
unconstrained generators $\hat{J}^{\bar{\A}}$ and the ghosts $c_\A$. 
We want to show that we can split the complex 
$\Gamma$ in a similar way:

\begin{prop}
\label{conj}
It is possible to 
define a subcomplex $\Gamma_{red}$ generated by modified 
unconstrained generators 
$\hat{W}^{\bar{A}}$ and ghosts $c_\A$,  
such that $\Gamma_{red}$ is a subcomplex (i.e. 
$s''(\Gamma_{red}) \subset \Gamma_{red})$. 
\end{prop} 

We will do the proof by a double induction, using the conformal 
dimension and the ($H-H'$)-grade of the generators as induction parameters. 

We consider the ``twisted'' algebra, i.e. the algebra where the conformal 
dimensions of generators without derivatives are given by the 
$H'$-grade + 1, and the $c$ ghosts have conformal dimension 0 
(alternatively we can think of it as simply a modified grade, 
defined by the $H'$-grade + 1 + the degree of derivatives). 
In this case the conformal 
dimensions of all the constrained generators is 1 and the 
($H-H'$)-grade of the constrained generators is less than zero. 

We will need a lemma: 
\begin{lem}
Consider an unconstrained generator $W^{\bar{\A}}$ with conformal dimension 
$h$ and grade $n$:
all unconstrained generators occurring in $s''(W^{\bar{\A}})$ has 
{\em either} conformal dimension strictly less than $h$ {\em or} 
conformal dimension 
$h$ and grade less than $n$. 
\end{lem}
In the expression $s''(W^{\bar{\A}})$, all generators  
 are the result of OPEs between a constrained generator 
in $j''$, and $W^{\bar{\A}}$. Thus all monomials\footnote{We use the word 
``monomial'', even though what we have is actually a normal-ordered product.} 
of generators 
occurring in $s''(W^{\bar{\A}})$ must have conformal dimension $h$ and 
$(H-H')$-grade less than $n$. We can write:
$$
s''(W^{\bar{\A}}) = P_{\bar{\B}}(c)W^{\bar{\B}} + 
Q_{\A\bar{\G}}(c) W^\A W^{\bar{\G}} + \cdots, 
$$
where $\cdots$ denote terms that are of higher order in the generators 
(constrained or unconstrained). 
The conformal dimension of $P_{\bar{\B}}(c)$ is greater than or equal to zero, 
so the conformal dimension of $W^{\bar{\B}}$ is less than or equal to $h$, 
and $(H-H')$-grade less than $n$. 
Similarly, the conformal dimension of $W^{\bar{\G}}$ is less than 
or equal to $h-1$, and 
we see that even stronger inequalities hold for the higher order terms. This  
proves the lemma. \\

Assume that we have already found hatted generators for all generators with 
conformal dimension less than $h$, 
and define $\Gamma_{red}^{h-1}$ to be the space generated 
by these hatted generators and the $c$'s. 
Assume that $W^{\bar{\A}}$ is any generator with conformal dimension $h$ and  
grade 0, we will show that we can define $\wh{W}^{\bar{\A}}$ such 
that $s''(\wh{W}^{\bar{\A}}) \in \Gamma_{red}^{h-1}$. 
Consider $s''(W^{\bar{\A}})$.
According to the lemma, all unconstrained generators 
occurring in $s''(W^{\bar{\A}})$ must have conformal dimension less than $h$. 
We can therefore write 
$$
s''(W^{\bar{\A}}) = \sum_{i,j} A_{ij} B_j,\quad A_{ij} \in 
\cb=\otimes_\B \Gamma^\B,\, 
B_j \in \Gamma_{red}^{h-1}
$$
where the $B_j$'s are chosen to be linearly independent. Since $j''$ is 
linear in 
the constrained currents,
each of the terms $A_{ij}$ are monomials in the constrained currents, the 
$W^\A$'s. Let us consider only those terms that have the highest grade, 
considered as monomials in $W^\A$. 
$$
s''(W^{\bar{\A}}) = \sum_{i,j} A^m_{ij} B_j + \mb{lower orders terms},\quad 
A^m_{ij} \mbox{ is order $m$ in $W^\A$} 
$$
Now apply $s''$ once again. We get:
$$ 
0 = \sum_{i,j} \left ( s''(A^m_{ij}) B_j \pm A^m_{ij} s''(B_j) \right ) 
$$
We know that $s''(B_j) \in \Gamma_{red}^{h-1}$, and 
$s''(A^m_{ij}) \in \cb$. We also know that $s''(A^m_{ij})$ is of order 
$m+1$ in the $W^\A$'s, and these are the only possible terms of order $m+1$;
and since the expression must vanish order by order in the $W^\A$'s, we find 
$$
0 = \sum_{i,j} s''(A^m_{ij}) B_j
$$
Since the $B_j$'s are linearly independent we find that 
$$
0 = \sum_i s''(A^m_{ij}) 
$$
Now we use the fact that $\cb$ has trivial cohomology: since 
$\sum_i A^m_{ij}$ is in the kernel of $s''$ it must be in the image of $s''$, 
so 
we can find $X_j$ (of grade $m-1$ in the $W^\A$'s) 
such that $s''(X_j) = \sum_i A^m_{ij}$. Define 
$$
W_{(1)} = W - \sum_j X_j B_j. 
$$
We find that: 
\beano
s''(W_{(1)}) & = &  \sum_{i,j} A^m_{ij} B_j + \mb{lower orders terms}
- \sum_j ( s''(X_j) B_j \pm  X_j s''(B_j) ) \\
& = & \sum_{i,j} A^m_{ij} B_j + \mb{lower order terms} 
- \sum_{i,j} A^m_{ij} B_j 
\mp \sum_{j} X_j s''(B_j) \\
& = & \mbox{lower order terms } \mp \sum_{j} X_j s''(B_j) 
\enano
(the $\pm$ depends on the Grassman parity of $X_j$). 
All these terms are of order at most $m-1$ in the $W^\A$'s. By induction we 
see that we can define $\hat{W}^{\bar{\A}}$ such that 
$s''(\hat{W}^{\bar{\A}})$ 
is a polynomial of 
degree 0 in the constrained currents. 

We want to show that in fact no $b$'s appear in $s''(\hat{W}^{\bar{\A}})$ 
either. 
Actually this is quite simple: write 
$$
s''(\hat{W}^{\bar{\A}}) = B + \sum_\A B_\A b^\A + \sum_{\A,\B} B_{\A\B} 
b^\A b^\B 
+ \cdots. 
$$
Apply $s''$ again to get 
$$
0 = s''(B) + \sum_\A s''(B_\A) b^\A \pm B_\A (\hat{J^\A}-\chi^\A) + \cdots 
$$
Since $s''(B_\A)$ does not contain any constrained currents, we must have
$0 = \sum_\A B_\A \hat{J}^\A$, but this can only be true if $B_\A=0$ for all 
$\A$. 
We see that indeed $s''(\hat{W}^{\bar{\A}}) \in \Gamma_{red}$. 

Now assume that we have found hatted generators for all generators with 
conformal dimension less than $h$, and with conformal dimension $h$ and 
grade less than $n$, 
and define $\Gamma_{red,n-1}^{h}$ to be the space generated 
by these hatted generators and the $c$'s. 
Assume that $W^{\bar{\A}}$ is any generator with conformal dimension $h$
and grade $n$, we want to show that we can define $\wh{W}^{\bar{\A}}$ such 
that $s''(\wh{W}^{\bar{\A}}) \in \Gamma_{red,n-1}^{h}$.
Consider $s''(W^{\bar{\A}})$.
According to the lemma, 
any unconstrained generator $W^{\bar{\B}}$ that occurs in 
$s''(W^{\bar{\A}})$ has {\em either} have conformal dimension less than $h$ 
{\em or} conformal dimension $h$ and grade less than $n$. 
We can therefore write 
$$
s''(W^{\bar{\A}}) = \sum_{i,j} A_{ij} B_j,\quad A_{ij} \in \cb,\, 
B_j \in \Gamma_{red,n-1}^{h}. 
$$
We can therefore repeat the arguments from above word by word to define 
$\hat{W}^{\bar{\A}}$ such that $s''(\hat{W}^{\bar{\A}}) \in \Gamma_{red}$.
 
We have shown that to any generator $W^{\bar{\A}}$ we can construct
$\hat{W}^{\bar{\A}}$ such that $s''(\hat{W}^{\bar{\A}}) \in \Gamma_{red}$. 
We have therefore shown that $\Gamma_{red}$ is a sub-complex. 

\rightline{$\Box$} 

\indent

Thus, $s''(\Gamma_{red}) \subset \Gamma_{red}$, and 
we can then use the K\"unneth theorem (see eq. (\ref{eq:kunneth}) to find 
\bea 
H^*(\Gamma ; s'' ) 
& \cong & H^* \left ( \Gamma_{red} \otimes (\bigotimes_\A \Gamma^\A ) ; s'' 
\right )\nn 
& \cong & H^* ( \Gamma_{red}; s'') \otimes \left 
( \otimes_\A H^*(\Gamma^\A;s'') 
\right ) \nn  
& \cong & H^* ( \Gamma_{red}; s'') 
\eea
Thus in order to calculate the cohomology $H^*(\Gamma ; s'' )$ it is in fact 
enough to calculate $H^* ( \Gamma_{red}; s'')$. 

Next step is to split $s''$ into two anti-commuting nilpotent operators 
$s''_0$ and $s''_1$ defined by the currents $j_0''$ and $j_1''$ respectively, 
\bea 
j_0''(z) & = & -\chi^\A c_\A(z) \nn
j_1''(z) & = & W^\A(z) c_\A(z) + \hf {f^{\A\B}}_\G b^\G(z) c_\B(z) c_\A(z)
\eea

In order to verify that $s''_0$ and $s''_1$ are indeed nilpotent and 
anti-commuting one can either directly calculate the operator products 
$j_1''(z) j_1''(w)$ etc., 
or one can use the fact that $s_0'' = s_0 - s_0'$ and $s_1'' = s_1 -
s_1'$, where $s_0,\,s_0',\,s_1$, and $s_1'$ are all nilpotent and
anti-commuting. 

Corresponding to this split, we can define a bigrading of $\Gamma$: 
\bea 
\label{eq:bigr}
W^\A,W^i & : & (m,-m)  \nn
b^\A     & : & (m,-m-1) \nn
c_\A     & : & (-m,m+1)
\eea 
where $m$ is the grade of $W^\A$ or $W^i$ defined by the grading $(H-H')$; 
with these definitions $s_0''$ has 
bigrading $(1,0)$, while $s_1''$ has bigrading $(0,1)$. 
We can now define the spectral 
sequence corresponding to the double complex $(\Gamma_{red} ; s_0'';s_1'')$.

The first element of the spectral sequence is 
$$
E_0^{p,q} = \Gamma^{p,q}_{red}, 
$$
while the second element is the cohomology 
of $s_0''$:
\ben
E_1^{p,q} \, = \, H^{p,q} (E_0 ; s_0'') 
\, = \, 
\frac{\Gamma_{red}^{p,q} \cap \ke(s_0'')}{\Gamma_{red}^{p,q} \cap \im(s_0'')}
\een

In the primary hamiltonian reduction one can show that for each ghost $c_A$,
we can find a linear combination of generators 
$a_{A\bar{A}}\hat{J}^{\bar{A}}$, 
such that 
$$
s_0(a_{A\bar{A}}\hat{J}^{\bar{A}}(z)) 
= s_0\left( a_{A\bar{A}} {f^{\bar{A}B}}_C (b^C c_B)_0(z)\right) 
= - a_{A\bar{A}} {f^{\bar{A}B}}_C \chi^C c_B(z) = c_A(z)
$$
If we replace the index $A$ by $\A$ in this equation, then since the index 
$\A$ 
corresponds to a generator with $H'$-grade zero and $\bar{A}$ has non-negative 
$H'$-grade, then we find that also $B$ and $C$ has $H'$-grade zero. 
This implies that $s_0'(a_{\A\bar{A}}\hat{J}^{\bar{A}})=0$, and since 
$s_0''=s_0-s_0'$ we find  
$$
s_0''(a_{\A\bar{A}}\hat{J}^{\bar{A}}(z)) = c_\A(z).
$$ 
This shows that the ghosts $c_\A$ are $s_0''$-exact, and the cohomology of 
$s_0''$ is only non-trivial at ghostnumber zero, i.e. we find 
\ben
\label{eq:gamma0}
E_1^{p,q} = H^{p,q} (E_0 ; s_0'') \cong \D_{p+q,0}\, \Gamma_0
\een
Where $\Gamma_0$ is the cohomology of $s_0''$ at ghostnumber zero. 

We recall that in the primary quantum hamiltonian reduction, the zeroth 
cohomology of $s_0$ is $\Omega_{hw}$. In the secondary hamiltonian reduction
there is no notion of highest weights, but $\Gamma_0$ can be considered to be 
the secondary hamiltonian reduction analogue of $\Omega_{hw}$.

Equation (\ref{eq:gamma0}), together with the fact that the bigrade of the 
operator $s_r$ is $(1-r,r)$ implies that $s_r$ is trivial for 
$r\geq 1$. 
Thus the spectral sequence collapses already here, and we have 
$$
E_\infty^{p,q} = E_1^{p,q} = H^{p,q} (E_0 ; s_0'') \cong \D_{p+q,0}\, \Gamma_0
$$

Using the equation (\ref{eq:th1}) of the theory of 
spectral sequences, we find that 
$$
F^q(H^{p+q})/F^{q+1}(H^{p+q}) = \D_{p+q,0}\, \Gamma_0
$$
where 
$F^q(H^{p+q})$ is defined as in eq. (\ref{toto}).

Note that with the bi-gradings defined in (\ref{eq:bigr}), 
$\Gamma_{red}^{p,q}$ is trivial for $q>0$, so 
$F^q \Gamma_{red}^{p+q} = \oplus_{i\geq 0} \Gamma_{red}^{p-i,q+i}$ is trivial 
for $q>0$, so we can use lemma \ref{eq:recH} page \pageref{eq:recH} and find 
\ben
\label{eq:cohs''}
H^n( \Gamma_{red} ; s'') \cong \D_{n,0}\,\Gamma_0;
\een
however, this isomorphism is a vector space isomorphism but not an algebra 
isomorphism. 

We have proven the following:
\begin{theo}
\label{the:expcoh}
The BRST operator $s''$ defined by 
\bea
s''(\phi)(w) & = & \oint_w dz\, j''(z) \phi(w) \nn
j''(z) & = & (W^\A(z)-\chi^\A)c_\A(z) + {f^{\A\B}}_\G (b^\G c_{\B}c_{\A})_0(z)
\eea 
corresponding to the secondary hamiltonian reduction 
$\cw(\cg,\ch')\rightarrow\cw(\cg,\ch)$ can be split into two anticommuting, 
nilpotent operators $s_0''$ and $s_1''$ defined by 
\beano
j_0''(z) & = &  -\chi^\A c_\A(z) \\
j_1''(z) & = &  W^\A(z) c_\A(z) + {f^{\A\B}}_\G (b^\G c_{\B}c_{\A})_0(z)
\enano 
The cohomology of $s''$ is 
$$
H^n( \Gamma_{red} ; s'') \cong \D_{n,0}\,\Gamma_0,
$$
where $\Gamma_0 = H^0(\Gamma_{red};s_0'')$. This isomorphism is a vector space
isomorphism, but not an algebra isomorphism. 
\end{theo}
Note that $\Gamma_{red}$ does not contain any elements of negative 
ghostnumber, and consequently $\Gamma_0 \cong \ke(s_0'')$. This, together 
with the
fact that we know the number of generators of $\Gamma_0$ (it is equal to 
the number of generators of \cw) considerably
simplifies the problem of finding $\Gamma_0$ in concrete examples. 

\newpage

\subsection{Explicit Construction of Generators.}

\label{sec:ttt}
Once we have found the generators $V_0^k$ of $\Gamma_0$, we can use the 
tic-tac-toe construction to find the generators of $H^0(\Gamma_{red},s'')$, 
i.e. the generators of \cw. 
These take the form 
$$
V^k(z) = \sum_{\ell=0}^p V^k_\ell(z)
$$
where $V^k_\ell$ is defined inductively by 
$s_1''(V^k_\ell) + s_0''(V^k_{\ell+1})=0$ and $p$ is given by 
$V^k_0 \in \Gamma_{red}^{p,-p}$ (i.e. it is the grade of $V^k_0$). 
Note that if $V^k_0$ has bigrade $(p,-p)$, $V^k_1$ has bigrade $(p-1,-p+1)$, 
etc. $V^k_p$ has bigrade $(0,0)$, and the construction stops here because 
$s_1''$ vanishes on generators with grade zero. 
It is easy to verify that with this construction, $s''(V^k)$ is indeed zero.  

The resulting generators $V^k$ constitutes a basis of the algebra \cw. 
In principle the operator product expansion of these generators close 
only modulo $s''$-exact 
terms; however, since there are no elements in $\Gamma_{red}$ with negative 
ghostnumber, there can be no $s''$-exact terms with zero ghostnumber, and 
therefore the algebra of the $V^k$ closes exactly. 

\subsection{Generalized Quantum Miura Transformation} 

Because the operator product expansion preserves the grading, it is clear that 
the grade zero part of the generators gives a copy of the \cw-algebra, or more 
precisely: the map $V^k \rightarrow V^k_p$ 
($V^k = V^k_0 + V^k_1 + \cdots + V^k_p$)
is an algebra homomorphism. In order to show that this map is in fact an
algebra isomorphism, we need to show that the map is an injection. 
To show this, one can consider the so-called ``mirror spectral sequence'', the 
spectral sequence obtained by inverting the role of $s_0''$ and $s_1''$. 
Thus for the mirror spectral sequence, we define 
\bea
\bar{E}_0^{p,q} & = & \Gamma_{red}^{p,q} \, (= E_0^{p,q}) \nn
\bar{E}_1^{p,q} & = & H^{p,q} (E_0 ; s_1'') 
\, = \, 
\frac{\Gamma_{red}^{p,q} \cap \ke(s_1'')}{\Gamma_{red}^{p,q} \cap \im(s_1'')}
\nn
\bar{E}_2^{p,q} & = & H^{p,q} (E_1 ; s_0'') 
\, = \, 
\frac{\bar{E}_1^{p,q} \cap \ke(s_0'')}{\bar{E}_1^{p,q} \cap \im(s_0'')}
\eea
etc. We already know that $H^*(\Gamma_{red};s'')$ is nontrivial only at 
ghostnumber zero. This implies that also $\bar{E}_\infty^{p,q}$ is nontrivial 
only at ghostnumber zero, i.e. at $q=-p$. 
We find that 
$s_1''(\hat{W}^{\bar{A}})=0$ iff $\hat{W}^{\bar{A}}$ has bi-grade $(0,0)$. 
To see this, note that $s_1''$ has bigrade (0,1), and that 
$\Gamma_{red}^{0,1}=\{ 0 \}$.
This shows that:
$$
\hat{W}^{\bar{A}} \mbox{ has bi-grade (0,0) } 
\Rightarrow s_1''(\hat{W}^{\bar{A}})=0.
$$ 
To see that the opposite is also true,
note that for each $W^{\bar{A}}$ with grade larger than zero, there is a 
$W^\A$ such that 
$$
\ope{W^\A}{W^{\bar{A}}} = \ord{(1+h_{\bar{A}})}{g^{\A\bar{A}}} + \cdots 
$$
where $\cdots$ denotes less singular terms. This gives rise to a 
term proportional to $\pa^{h_{\bar{A}}} c_{\A}$ in $s_1''(\hat{W}^{\bar{A}})$; 
and one can show that this term will not be cancelled by other terms in 
$s_1''(\hat{W}^{\bar{A}})$,
thus showing that 
$s_1''(\hat{W}^{\bar{A}})\not = 0$.

It follows that 
$$ 
\bar{E}_1^{p,-p} = \D_{p,0}\,\Gamma_{red}^{0,0}
$$
Therefore there is an isomorphism of vector spaces 
$$
H^0(\Gamma_{red};s'') \cong \bar{E}_\infty^{0,0} 
$$ 
and therefore the map from $H^0(\Gamma_{red};s'')$ to its zero grade component 
is injective, and therefore indeed an isomorphism of algebras. 
This proof is essentially identical 
to the one given in \cite{BoTj} for the case of the primary hamiltonian 
reduction. 

We have shown: 
\begin{theo}
\label{the:miura}
For generators $V^k$ of \cw, constructed using the tic-tac-toe construction 
defined above, the mapping 
$$
V^k = V^k_0 + V^k_1 + \cdots + V^k_p \rightarrow V^k_p
$$ 
of the generator to the zero grade part of the generator is an algebra 
isomorphism. 
\end{theo}

This mapping is the generalization of the quantum Miura transformation 
to the case of the secondary hamiltonian reduction. 

This theorem means that we can realize the generators of the algebra \cw~
in terms of the generators of the simpler algebra $\hat{\cw}_0'$, 
generated by the ``hatted'' grade zero generators of $\cw'$. $\hat{\cw}_0'$ 
always includes the energy-momentum tensor $\hat{T}$, since $T$ is 
always part of the grade zero subspace of $\cw'$. 

This construction gives us an impressive variety of new realizations of \cw~
algebras: for every possible secondary hamiltonian reduction, written in the 
form 
$$
\cg^{(1)} \rightarrow \cw(\cg,\ch') \rightarrow \cw(\cg,\ch),
$$
we get a realization of the generators of $\cw(\cg,\ch)$ in terms of 
the hatted generators of the grade zero subalgebra of $\cw(\cg,\ch')$.

Similar realizations of \cw~algebras in terms of simpler \cw~algebras 
have been constructed before, see e.g. 
\cite{DaDhRa,De}; however, the present construction gives a systematic 
method for constructing a large number of such realizations. 

\subsection{Example: $\cw(s\ell(3),s\ell(2)) \rightarrow \cw_3$}
\label{exa:1}

Let us consider the simplest possible example of the secondary quantum
hamiltonian reduction, namely the reduction of the Bershadsky algebra 
$\cw(s\ell(3),s\ell(2))$ to the $\cw_3$ algebra. 
We consider the regular embedded $s\ell(2)$ subalgebra 
$\{ E_{\A_1} , H_{\A_1} , E_{-\A_1} \}$. The corresponding standard grading of
$s\ell(3)$ is 
$$
\left ( \begin{array}{rrr}
0    & 1   &  \hf \\
-1   & 0   & -\hf \\
-\hf & \hf &  0   
\end{array} \right )  
$$
As described in section \ref{sec:primared}, we can modify the grading 
operator $H_{\A_1}$ by adding a $U(1)$ current. If we choose 
$$
U = \left ( \begin{array}{rrr} \frac{1}{6} & 0 & 0 \\
0 & \frac{1}{6} & 0 \\ 0 & 0 & - \frac{1}{3} \end{array} \right ) 
$$
then we get the modified integer gradings 
$$
\left ( \begin{array}{rrr}
0  & 1   &  1 \\
-1 & 0   &  0 \\
-1 & 0 &  0   
\end{array} \right )  
$$
The constrained current corresponding to these gradings are 
\ben 
\label{eq:con'1}
J_{red} = \left ( \begin{array}{ccc}
H^{\A_1} & J^{\A_1}          &  J^{\A_1+\A_2} \\
  1      & H^{\A_2}-H^{\A_1} & J^{\A_2}       \\
  0      & J^{-\A_2}         &  -H^{\A_2}    
\end{array} \right )  
\een
(to simplify the notation we suppress the $z$ dependence). 
The grading and constraints for the $\cw_3$--algebra are: 
\ben
\label{eq:con1}
\left ( \begin{array}{rrr}
0  &  1  &  2 \\
-1 &  0  &  1 \\
-2 & -1  &  0   
\end{array} \right )  
\quad \quad 
J_{red} = \left ( \begin{array}{ccc}
H^{\A_1} & J^{\A_1}          &  J^{\A_1+\A_2} \\
  1      & H^{\A_2}-H^{\A_1} & J^{\A_2}       \\
  0      &     1             &  -H^{\A_2}    
\end{array} \right )  
\een
We introduce ghosts $c_{-\A_1},\,c_{-\A_2},\,c_{-\A_1-\A_2},\,$ and 
anti-ghosts 
$b^{-\A_1},\,b^{-\A_2},\,b^{-\A_1-\A_2}$. Corresponding to the constraints 
(\ref{eq:con'1}) and (\ref{eq:con1}) we have the BRST currents $j'$ and $j$
respectively: 
\bea
j' & = & (J^{-\A_1} - 1)c_{-\A_1} + J^{-\A_1-\A_2} c_{-\A_1-\A_2} \\
j  & = & (J^{-\A_1} - 1)c_{-\A_1} + (J^{-\A_2} - 1)c_{-\A_2} 
+ J^{-\A_1-\A_2} c_{-\A_1-\A_2} + b^{-\A_1-\A_2} c_{-\A_1} c_{-\A_2} \nonumber
\eea
We define ``improved'' generators 
$$
\begin{array}{rclrcl}
\tilde{J}^{-\A_1} & = & J^{-\A_1} & 
\tilde{J}^{\A_1} & = & J^{\A_1} \\
\tilde{J}^{-\A_2} & = & J^{-\A_2} + b^{-\A_1-\A_2} c_{-\A_1} &
\tilde{J}^{\A_2} & = & J^{\A_2} + b^{-\A_1} c_{-\A_1-\A_2} \\
\tilde{J}^{-\A_1-\A_2} & = & J^{-\A_1-\A_2} & 
\tilde{J}^{\A_1+\A_2} & = & J^{\A_1+\A_2}                \\
\tilde{H}^{\A_1} & = & H^{\A_1} - 2 b^{-\A_1} c_{-\A_1} 
- b^{-\A_1-\A_2} c_{-\A_1-\A_2} \\
\tilde{H}^{\A_2} & = & H^{\A_2} + b^{-\A_1} c_{-\A_1} 
- b^{-\A_1-\A_2} c_{-\A_1-\A_2} 
\end{array}
$$
and we find that 
$\cw(s\ell(3),s\ell(2)) = H^0(\ca\otimes\LM';s')$ is generated by 
\bea
J & = & \tilde{H}^{\A_1} + 2 \tilde{H}^{\A_2} \\
G^- & = & \tilde{J}^{-\A_2} \nn
G^+ & = & \tilde{J}^{\A_1+\A_2} +(k+2) \pa \tilde{J}^{\A_2} -
(\tilde{H}^{\A_1} \tilde{J}^{\A_2})_0 - (\tilde{H}^{\A_2} \tilde{J}^{\A_2})_0
\nn
T  & = & \frac{1}{k+3} \left [ \tilde{J}^{\A_1} 
- \frac{1+k}{2} \pa\tilde{H}^{\A_1}  
+ ( \tilde{J}^{\A_2} \tilde{J}^{-\A_2})_0 
+ \th \left ( (\tilde{H}^{\A_1} \tilde{H}^{\A_1})_0 
+ (\tilde{H}^{\A_1} \tilde{H}^{\A_2})_0 
+ (\tilde{H}^{\A_2} \tilde{H}^{\A_2})_0 \right ) \right ]. \nonumber
\eea
Here J is a $U(1)$ field and $G^\pm$ are primary bosonic spin $\thf$ fields. 
With the normalizations chosen, we have 
\bea 
\ope{J}{J} & = & \ord{2}{9+6k} +\cdots \\
\ope{J}{G^\pm} & = & \ordo{\pm 3 G^\pm(w)} +\cdots \nn
\ope{G^+}{G^-} & = & -\ord{3}{(k+1)(2k+3)} - \ord{2}{(k+1)J(w)} 
+ \ordo{(k+3) T - \frac{k+1}{2} \pa J - \th (JJ)_0}  +\cdots  \nonumber
\eea 
where $\cdots$ denotes non-singular terms. 
The central charge is $c = - \frac{(2k+3)(3k+1)}{(k+3)}$. 

The BRST current for the secondary hamiltonian reduction can now be written 
in the form:
\ben 
j'' = (J^{-\A_2} - 1)c_{-\A_2} + b^{-\A_1-\A_2} c_{-\A_1} c_{-\A_2} 
= (G^- - 1)c_{-\A_2} 
\een
The operator $s''$ is found to act on the fields as follows: 
\bea 
s''(T) & = & \frac{3}{2} G^- c_{-\A_2} 
+ \hf \pa G^- c_{-\A_2}  \nn
s''(J) & = & 3 G^- c_{-\A_2} \nn
s''(G^+) & = & -(k+3) T c_{-\A_2} - (k+1) J \pa c_{-\A_2} 
- \frac{k+1}{2} \pa J c_{-\A_2} + \th (JJc_{-\A_2})_0 \nn
s''(b^{-\A_2}) & = & G^- - 1
\eea
and $s''(G^-) = s''(c_{-\A_2}) = 0$. The ``hatted'' operators are: 
\bea 
\hat{T} & = & T - \frac{3}{2} b^{-\A_2} \pa c_{-\A_2} + 
\hf c_{-\A_2} \pa b^{-\A_2} \nn
\hat{G}^- & = & G^- \nn
\hat{G}^+ & = & G^+ \nn
\hat{J} & = & J - 3 b^{-\A_2} c_{-\A_2} 
\eea
and we find that $s''(\hat{T}) = \frac{3}{2} \pa c_{-\A_2}$ and 
$s''(\hat{J}) = 3 c_{-\A_2}$. 
The operators $s_0''$ and $s_1''$ are given in terms of the currents 
$j_0''$ and $j_1''$ respectively: 
\beano
j_0'' & = & -c_{-\A_2} \\
j_1'' & = & G^- c_{-\A_2} 
\enano
We find that the generators of $\Gamma_0 = H^0(\Gamma_{red} ;s_0'')$
are $T_2 = \hat{T} - \hf \pa\hat{J}$ and $G^+$. $T_2$ is already in the 
cohomology of $s''$, and using the tic-tac-toe construction with $G^+$ 
as the starting point, we find $W$: 
\bea
\label{eq:tw}
T_2 & = & \hat{T} - \hf \pa\hat{J} \\
W & = & G^+ - \frac{1}{27} (\hat{J}\hat{J}\hat{J})_0 
+ \frac{1+k}{6} (\hat{J}\pa\hat{J})_0 + \frac{(k+3)}{3} (\hat{T}\hat{J})_0 \nn
& & - \frac{(k+3)(k+2)}{2} \pa \hat{T} - \frac{(k+3)k+4}{12}\pa^2 \hat{J} 
\nonumber
\eea
This gives us a realization of the $\cw_3$ algebra in terms of the generators 
of $(\hat{\cw}_3^2)_{\geq 0}$, the ``hatted'' generators of $\cw_3^2$ with 
non-negative grade. 

Using the primary hamiltonian reduction, we can find expressions for  
$G^+$, $T$, and 
$J$ in terms of the currents of the affine algebra $s\ell(3)^{(1)}$. 
Inserting these expressions into equation (\ref{eq:tw}) gives us a 
realization of $\cw_3$ in terms of the currents of the $s\ell(3)^{(1)}$.
Note, however, that this is not identical to the realization we would get 
by doing the hamiltonian reduction to $\cw_3$ in one step, using 
the primary hamiltonian reduction. 

\sect{Linearization of \cw-algebras}
\label{sec:lin}

Very recently, the construction of linearized \cw~algebras \cite{BeKrSo} have 
attracted some attention. The idea in this construction is
to add some extra generators to an algebra \cw, such that the resulting 
larger algebra is equivalent to a linear algebra. 

We will show that the secondary quantum hamiltonian reduction gives us a 
general method to find such linearizations of \cw~ algebras. In 
the specific case of the linearization of $\cw_3$, we find the same result as 
\cite{BeKrSo}. 

The basic idea of our construction is very simple. Define $\cw_-'$ to be the
subalgebra of $\cw'$ with negative grading, i.e. the  
constrained subalgebra of $\cw'$, and define $\cw'_{\geq 0}$ to be the 
subalgebra with nonnegative grading. Define furthermore $\hat{\cw}_{\geq 0}'$
to be the algebra generated by the ``hatted'' generators in $\cw_{\geq 0}$. 
We have above shown (property \ref{conj}) that we can construct a 
realization of \cw~as differential 
polynomials in the generators of $\hat{\cw}_{\geq 0}'$. 
Let us denote the number of generators of an algebra\footnote{It remains to
show that \ca\ is indeed an algebra.} \ca~by $|\ca|$, and 
define 
$$
n = |\cw| -  |\hat{\cw}_{\geq 0}'| = |\cw_-'|. 
$$
We will show that it is possible to add $n$ of the generators of 
$\hat{\cw}_{\geq 0}'$ to $\cw$, in such a way that there is a invertible 
transformation between the resulting algebra $\cw_{ext}$ and 
$\hat{\cw}_{\geq 0}'$. 

Let us write $\hat{\cw}_{\geq 0}'$ in the form 
$$
\hat{\cw}_{\geq 0}' = \Gamma_0 \oplus V.  
$$ 
($V$ is not uniquely defined). 
It follows from the tic-tac-toe construction that the generators of 
\cw~have the form: 
$$
W = W_0 + W_1, \quad W_0 \in \Gamma_0,\, W_1 \in V.
$$
It is now clear, that if we extend \cw~with a basis of generators in $V$, 
then the 
transformation $\hat{\cw}_{\geq 0}' \leftrightarrow \cw_{ext}$ is invertible. 
We have 
\begin{theo} 
Write the algebra $\hat{\cw}_{\geq 0}'$ in the form 
$\hat{\cw}_{\geq 0}' = \Gamma_0 \oplus V$, where $\Gamma_0 = \ke(s'')$. 
Define $\cw_{ext}$ to be the algebra 
\cw extended with a basis of generators in $V$. Then there is an 
invertible mapping 
$$
\phi:\hat{\cw}_{\geq 0}' \rightarrow \cw_{ext}
$$
\end{theo} 

We see that every secondary hamiltonian reduction gives rise to an embedding 
$\cw \hookrightarrow \cw_{ext}$, where $\cw_{ext}$ is equivalent to an algebra 
$\hat{\cw}_{\geq 0}'$ that will in general be simpler than \cw. 

If $\hat{\cw}_{\geq 0}'$ is linear, the result of this procedure is 
a linearization of \cw. Generically, $\hat{\cw}_{\geq 0}'$ is not linear,
but we find that it is actually linear for a large class of reductions:
\begin{prop}
All algebras of the form 
$$
\cw(s\ell(N),\oplus_{n=1}^l s\ell(p_n)\ ),\quad p_1>p_n+1, \quad 
\forall\ n\geq2
$$
can be linearized by the secondary hamiltonian reduction
$$
\cw(s\ell(N), s\ell(2)) \rightarrow 
\cw(s\ell(N),\oplus_{n=1}^l s\ell(p_n)\ ). 
$$
The tic-tac-toe construction gives an algorithmic method for the explicit 
construction of these linearizations. 
\end{prop} 
Let us restrict ourselves to showing this in the case of 
$\cw(s\ell(n),s\ell(m))$ -- the general case is a 
straightforward generalization. 
So we consider the secondary reduction 
$\cw(s\ell(n),s\ell(2)) \rightarrow \cw(s\ell(n),s\ell(m))$. The constraints 
and highest weight gauge corresponding to $\cw(s\ell(n),s\ell(2))$ are 
$$
\left ( \begin{array}{ccccc}
     * & * & \cdots & * & * \\ 
     1 & * & \cdots & * & * \\
     0 & * & \cdots & * & * \\
\vdots & \vdots & & \vdots & \vdots \\ 
     0 & * & \cdots & * & *
\end{array} \right ) 
\quad \quad \quad 
\left ( \begin{array}{c|c} 
\begin{array}{cc} U & T \\ 1 & U 
\end{array}  & 
\begin{array}{cccc} G_1 & G_2 & \cdots & G_{n-2} \\ 0 & 0 & \cdots & 0 
\end{array} \\ \hline
\begin{array}{cc} 
0 & \bar{G}_1 \\
0 & \bar{G}_2 \\
\vdots & \vdots \\
0 & \bar{G}_{n-2} 
\end{array} & 
s\ell(n-2) + \frac{2U}{n-2} \1
\end{array} \right )
$$
The $G$'s are bosonic spin $\thf$ fields. 
The $U(1)$ operator $U$ commutes with the $s\ell(n-2)$ Kac-Moody subalgebra 
in $\cw(s\ell(n),s\ell(2))$, while the $G$'s have positive and the $\bar{G}$'s
negative $U(1)$-charge. This shows that the only possible operator product 
expansions containing nonlinear terms are $G_i(z) \bar{G}_j(w)$. 
The constraints corresponding to $\cw(s\ell(n),s\ell(m))$ are 
$$
\left ( \begin{array}{ccccccc} 
     * &      * & \cdots & * &      * & \cdots & * \\ 
     1 &      * & \cdots & * &      * & \cdots & * \\
     0 &      1 & \cdots & * &      * & \cdots & * \\
\vdots & \vdots & \ddots &   & \vdots &        & \vdots \\
     0 &      0 & \cdots &      1 &      * & \cdots & * \\
     0 &      0 & \cdots &      0 &      * & \cdots & * \\
\vdots & \vdots &        & \vdots & \vdots &        & \vdots \\
     0 &      0 & \cdots &      0 &      * & \cdots & * 
\end{array} \right )        
$$
We find that the secondary reduction is made by constraining 
$\bar{G}_1 = 1, \, \bar{G}_2 = 0, \ldots, \bar{G}_{n-2} = 0$, in general in 
addition to constraining also a number of the Kac-Moody currents. Since all 
the fields $\bar{G}_i$ are constrained, it follows that $\hat{\cw}_{\geq 0}$ 
is linear. 

For the $so(n)$ algebras, due to the few cases that allow the secondary 
reductions (in our 
framework), it is clear that we will not be able to linearize most of the 
corresponding 
\cw-algebras. In fact, demanding that the starting \cw-algebra is built on 
$\ch=s\ell(2)$ and 
reasoning as above, it is easy to see that only the algebras\footnote{We 
denote by $\cw 
BC_n$ the algebra $\cw(B_n,B_n)$ obtained from Hamiltonian reduction of 
$B_n$ to 
distinguish it from the Casimir algebra $\cw B_n$ that contains a spin 
$\frac{n+1}{2}$ field. 
$\cw BC_n$ as the same spin contents as $\cw C_n$ but different structure 
constants.} 
 $\cw BC_2$ and $\cw D_3$ can be linearized (the last one being in fact 
identical with 
$\cw A_3$).

For $sp(2n)$ algebras, the complete classification of linearizable 
$\cw(sp(2n),\ch)$ algebras is quite
heavy and beyond the scope of the present article: we refer to 
\cite{MaRa2} for an exhaustive
classification. Let us just remark that the secondary reduction 
$\cw(sp(2n),sp(2)) \rightarrow \cw(sp(2n),\ch)$ with \ch\ simple always 
provide a linearization of
the $\cw(sp(2n),\ch)$ algebra.

Let us remark that the spin of the new fields we add to linearize the 
algebra are always positive, 
since we take the positive grade part of a given 
\cw-algebra\footnote{Actually the spin is at least
1.}.

The most popular \cw-algebras are the $\cw(\cg, \cg)\equiv\cw\cg$ ones: 
it is natural to see whether 
one can linearize these algebras. From the above property, it is easy to 
deduce:
\begin{prop}
The \cw-algebras \cw$A_n$ and \cw$C_n$ can be linearized by the secondary 
reductions 
through the schemes:
$$
\begin{array}{l}
\cw\left(s\ell(n+1), s\ell(2) \right) \rightarrow \cw A_n \\
\cw\left(sp(2n), sp(2) \right) \rightarrow \cw C_n 
\end{array}
$$
For the \cw$BC_n$, and \cw$D_n$ algebras, our techniques allows to linearize 
only the 
\cw$BC_2$ and \cw$D_3$ algebras through
$$
\begin{array}{l}
\cw\left(so(5), so(3)\right) \rightarrow \cw BC_2 \\
\cw\left(so(6), so(3)\right) \rightarrow \cw D_3
\end{array}
$$
\end{prop}

The above method of linearizing \cw~algebras is not limited to the 
secondary quantum hamiltonian reduction -- it can also be used in the primary 
quantum hamiltonian reduction. Following the above procedure in that case, we 
find any \cw~algebra $\cw(\cg,\ch)$ can be extended by adding the generators 
in $\hat{\cg}^{(1)}_{\geq 0}$ which are {\em not} highest weight, 
$\hat{\cg}^{(1)}_{\geq 0}$ are the hatted affine currents with non-negative 
grade:
$$
\cw_{ext} = \cw + \left (\hat{\cg}^{(1)}_{\geq 0} 
\setminus \hat{\cg}^{(1)}_{hw} \right ) 
$$
The algebra $\cw(\cg,\ch)_{ext}$ is then equivalent to 
$\hat{\cg}^{(1)}_{\geq 0}$. 

As explicit examples of the linearization using secondary hamiltonian 
reduction, we will give the two simplest: 
the linearization of $\cw_3 = \cw(s\ell(3),s\ell(3))$ using 
$(\widehat{\cw}_3^2)_{\geq 0}  = \widehat{\cw}(s\ell(3),s\ell(2))_{\geq 0}$ 
(this linearization was already given in \cite{BeKrSo}), and the linearization 
of $\cw_4 = \cw(s\ell(4),s\ell(4))$ using 
$(\widehat{\cw}(s\ell(3),s\ell(2))_{\geq 0}$.

Note also that the secondary reduction 
$\cw(so(5),so(3)) \rightarrow \cw(so(5))$ 
will provide the linearization of the $W_{2,4}$ algebra \cite{BeKrSo}. For the 
linearization of the 
$\cw B_2$ algebra (containing a spin $\frac{5}{2}$ field), a secondary 
reduction of super algebras will have to be performed \cite{MaRa2}. 

\subsection{Linearization of $\cw_3$}
As we already saw in example in section \ref{exa:1}, $\cw_3$ can be realized 
in terms of the generators $\hat{T},\, \hat{G}^+$, and $\hat{J}$: 
\bea 
T & = & \hat{T} - \hf \pa\hat{J} \\
W & = & \hat{G}^+ - \frac{1}{27} (\hat{J}\hat{J}\hat{J})_0 
+ \frac{1+k}{6} (\hat{J}\pa\hat{J})_0 + \frac{(k+3)}{3} (\hat{T}\hat{J})_0 \nn 
& & - \frac{(k+3)(k+2)}{2} \pa \hat{T} - \frac{(k+3)k+4}{12}\pa^2 \hat{J} 
\eea
If we add the current $J=\hat{J}$ to the $\cw_3$ algebra, then it is clear 
that the transformation 
$\{ T,W,J \} \leftrightarrow \{ \hat{T}, \hat{G}^+,\hat{J} \}$ is invertible. 
The new operator product expansions of the extended $\cw_3$ algebra are: 
\bea 
\ope{J}{J} & = & \ord{2}{18+6k} +\cdots \nn
\ope{T}{J} & = & \ord{3}{12+6k} + \ord{2}{J} + \ordo{\pa J} +\cdots \nn
\ope{J}{W} & = & \ord{3}{(k^2+5k+6)J} 
+ \ord{2}{(2k^2+12k+18)T - \frac{1}{3}(k+3)(JJ)_0 + \hf(3k^2+15k+18)\pa J} \nn
& & + \ordo{3 W + \frac{3}{2}(k^2+5k+6)\pa T + \hf(2k^2+9k+11)\pa^2 J 
-(k+3)(TJ)_0} \nn 
& & + \ordo{\frac{1}{9}(JJJ)_0 - (k+2) (J\pa J)_0 } +\cdots 
\eea
while the (equivalent) nontrivial operator product expansions of the 
linear algebra generated by 
$\hat{T},\, \hat{G}^+$, and $\hat{J}$ are 
\bea 
\ope{\hat{T}}{\hat{T}} & = & - \ord{4}{\frac{3k^2+11k+18}{k+3}} 
+ \ord{2}{2 \hat{T}} + \ordo{ \pa \hat{T}} + \cdots \nn
\ope{\hat{T}}{\hat{G}^+} & = & \ord{2}{\frac{3}{2} \hat{G}^+} 
+ \ordo{\pa\hat{G}^+} + \cdots \nn
\ope{\hat{T}}{\hat{J}} & = & \ord{3}{-6} + \ord{2}{\hat{J}} 
+ \ordo{\pa \hat{J}} + \cdots \nn
\ope{\hat{J}}{\hat{G}^+} & = & \ordo{3 \hat{G}^+ } + \cdots 
\eea 
and of course $\ope{\hat{J}}{\hat{J}} = \ope{J}{J}$.

\subsection{Linearization of $\cw_4$}

In order to show an example where the linearization has not been done before, 
we take the linearization of the $\cw_4$-algebra. In this case, the algebra 
$\cw(s\ell(4),s\ell(2))$ contains $T$, a $U(1)$ subalgebra generated by $U$, 
an affine $s\ell(2)$ algebra generated by $K^0$ and $K^\pm$, and 4 spin 
$\frac{3}{2}$ fields $G^{\epsilon\sigma},\,\epsilon = \pm,\,\sigma=\pm$. 
$G^{\epsilon\sigma}$ has $U(1)$-charge $\epsilon 1$ and the eigenvalue under 
$K^0$ is $\sigma \hf$. 

In the secondary reduction we constrain $G^{-\pm}$ and $K^-$, so the algebra
$\hat{\cw}(s\ell(4),s\ell(2))_{\geq 0}$ is generated by 
$\hat{T},\,\hat{G}^{+\pm},\,\hat{K}^0,\,\hat{K}^+$, and $\hat{U}$. 
$\hat{G}^{+\pm}$ are primary Virasoro and Kac-Moody fields, with 
spin $\thf$, $U(1)$-charge 1, and eigenvalue $\pm\hf$ 
under $\hat{K}^0$. $\hat{K}^+$ is a primary spin 1 field with eigenvalue 
$1$ under $\hat{K}^0$ (and $U(1)$-charge 0). the central charge is 
$\hat{c} = - \frac{3(2k^2+11k+32)}{k+4}$, and the rest of the nontrivial   
operator product expansions are: 
\bea 
\ope{\hat{T}}{\hat{U}} & = &  \ord{3}{-4} + \ord{2}{\hat{U}} + 
\ordo{\pa\hat{U}} +\cdots \nn
\ope{\hat{T}}{\hat{K}^0} & = &  \ord{3}{-1} + \ord{2}{\hat{K}^0} + 
\ordo{\pa\hat{K}^0} + \cdots \nn
\ope{\hat{U}}{\hat{U}} & = & \ord{2}{k+4} + \cdots \nn
\ope{\hat{K}^0}{\hat{K}^0} & = & \ord{2}{\frac{k+4}{2}} + \cdots \nn
\ope{\hat{K}^+}{\hat{G}^{+-}} & = & \ordo{-\hat{G}^{++}} + \cdots 
\eea

The tic-tac-toe construction gives us the expressions for the generators of 
$\cw_4$: 
\bea
T & = & \hat{T} - \pa \hat{K}^0 - 2 \pa \hat{U} \nn
W_3 & = & \hat{G}^{+-} - 2 (\hat{K}^+\hat{U})_0 + (2k+6) \pa\hat{K}^+   
 + (4+k)(\hat{T}\hat{U})_0  -\hf (\hat{U}\hat{U}\hat{U})_0 \nn
& & -2 (\hat{U}\hat{K}^0\hat{K}^0)_0 + (k+1) (\hat{U}\pa\hat{U}')_0 + 
   (k+2) (\hat{U}\pa\hat{K}^0)_0 + 4(k+3) (\hat{K}^0\pa\hat{K}^0)_0 \nn
& &  + \frac{(3k+8)}{2} \pa^2\hat{U} - (k+2)(k+3) \pa^2\hat{K}^0 -
(k+3)(k+4)) \pa \hat{T} \nn
W_4 & = & \hat{G}^{++} + (\hat{K}^+\hat{K}^+)_0 + \hf (\hat{G}^{+-}\hat{U})_0 
+ (\hat{G}^{+-}\hat{K}^0)_0 - (k+4) (\hat{T}\hat{K}^+)_0  \nn
& & + \hf (\hat{U}\hat{U}\hat{K}^+)_0  
+ 2 (\hat{K}^0\hat{K}^0\hat{K}^+)_0 - (k+3) (\hat{U}\pa\hat{K}^+)_0 \nn 
& & - (k+1)(\pa\hat{U}\hat{K}^+)_0 -2 (\hat{K}^0\pa\hat{K}^+)_0 
-k (\pa\hat{K}^0\hat{K}^+)_0 + \frac{(22+13k+2k^2)}{2} \pa^2\hat{K}^+  \nn
& & + \frac{(4+k)^2(952+643k+108k^2)}{4(2552+1763k+300k^2)} 
 (\hat{T}\hat{T})_0 + \frac{k+4}{4} (\hat{T}\hat{U}\hat{U})_0 
- (k+4) (\hat{T}\hat{K}^0\hat{K}^0)_0  \nn
& & - \frac{3}{16} (\hat{U}\hat{U}\hat{U}\hat{U})_0 
+ \hf (\hat{U}\hat{U}\hat{K}^0\hat{K}^0)_0 
+ (\hat{K}^0\hat{K}^0\hat{K}^0\hat{K}^0)_0 
+ \frac{k+1}{4} (\hat{U}\hat{U}\pa\hat{U})_0 
\nn
& & 
+ \frac{(4+k)(40+793k+513k^2+84k^3}{2(2552+1763k+300k^2)}(\hat{T}\pa\hat{U})_0
- \frac{3k+10}{4} (\hat{U}\hat{U}\pa\hat{K}^0)_0 \nn
& & + \frac{(4+k)(11504+12158k+4251k^2+492k^3)}{2(2552+1763k+300k^2)} 
(\hat{T}\pa\hat{K}^0)_0
- 2(3+k) (\hat{U}\hat{K}^0\pa\hat{K}^0)_0  \nn
& &
- (k+2) (\hat{K}^0\hat{K}^0\pa\hat{K}^0)_0 
- \frac{3(4+k)(8+3k)(13+4k)(184+121k+20k^2)}{8(2552+1763k+300k^2)} 
 \pa^2\hat{T} \nn
& & 
+ \frac{7336+9073k+4814k^2+1265k^3+132k^4}{4(2552+1763k+300k^2)} 
(\pa\hat{U}\pa\hat{U})_0 + (k^2+6k+9)(\hat{U}\pa^2\hat{K}^0)_0 \nn
&& + \frac{15152+27782k+18163k^2+5077k^3+516k^4}{2(2552+1763k+300k^2)}
 (\pa\hat{U}\pa\hat{K}^0)_0 \nn
& &
+ \frac{168352+235972k+123812k^2+28811k^3+2508k^4}{4(2552+1763k+300k^2)} 
 (\pa\hat{K}^0\pa\hat{K}^0)_0 \nn 
& & -(k+1) (\pa\hat{U}\hat{K}^0\hat{K}^0)_0  
+ (2k^2+13k+22)(\hat{K}^0\pa^2\hat{K}^0)_0 
-\frac{k+4}{4} (\hat{U}\pa^2\hat{U})_0 \nn  
& & + 
\frac{244688+354290k+201124k^2+55477k^3+7326k^4+360k^5}{12(2552+1763k+300k^2)}
\pa^3\hat{U} \nn 
 && -
\frac{1179328+1976920k+1325876k^2+445043k^3+74808k^4+5040k^5}
{24(2552+1763k+300k^2)} 
\pa^3\hat{K}^0 \nn
& & - \frac{k+3}{2} \pa W_3
\eea
We define the algebra $(\cw_4)_{ext}$ by 
adding the generators $\hat{K}^+,\hat{K}^0$, and $\hat{U}$ to the $\cw_4$ 
algebra. It is obvious that there is and invertible transformation between 
this extended algebra, and the linear algebra generated by 
$\hat{T},\,\hat{G}^{++}\,\hat{G}^{+-}\,\,\hat{K}^+,\hat{K}^0$, and $\hat{U}$.


\sect{Conclusion}

In this paper, we have considered secondary quantum hamiltonian reductions, 
i.e. hamiltonian reductions that can be described by the diagram: \\
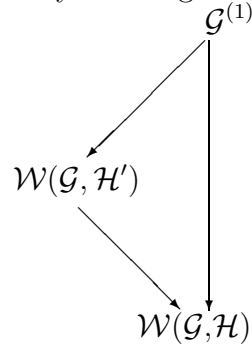
\begin{picture}(500,150)(-200,-140)
\setlength{\unitlength}{0.3mm}
\put(40,0){$\cg^{(1)}$}
\put(40,-4){\vector(-1,-1){52}}
\put(-45,-70) {$\cw(\cg,\ch')$}
\put(42,-4) {\vector(0,-1){120}} 
\put(10,-135) {\cw(\cg,\ch)} 
\put(-16,-78) {\vector(1,-1){45}}
\end{picture}\\
or in words: starting with a Lie algebra \cg, and two regular subalgebras 
$\ch'$ and $\ch$ with $\ch'\subset\ch$ satisfying certain conditions as
described in appendix \ref{sec:u1}, we carry out the hamiltonian reduction 
of the \cw~algebra $\cw(\cg,\ch')$ with suitable constraints, and show that 
the result is the \cw~algebra \cw(\cg,\ch). 

Note that for \cg=$s\ell(N)$, the conditions that we impose on $\ch'$ and 
$\ch$ in order to
perform the secondary quantum hamiltonian reduction are more restrictive 
than the conditions necessary for the classical secondary hamiltonian 
reduction, see \cite{DeFrRaSo}. This should not be taken as a sign that 
not all classical secondary hamiltonian reductions can be quantized; it simply 
reflects the fact that the method that we have used for the quantum secondary 
reduction in this paper cannot be applied to all possible secondary 
reductions. 
 On the other hand, we have been able to explicitly do some quantum 
reductions when 
 $\cg=so(N)$ or $\cg=sp(2N)$, while the techniques has not been 
developped for the classical case.

The quantum secondary reductions show that the \cw~algebras 
\cw(\cg,\ch) that 
can be obtained by the hamiltonian reduction of a certain affine Lie algebra 
$\cg^{(1)}$ are not only related by their common ``ancestor'' $\cg^{(1)}$, 
but that they are mutually directly connected by the hamiltonian reduction. 
As a simple example, consider this diagram of the possible hamiltonian 
reductions connecting the algebras $\cw(s\ell(4),\ch)$; the simple lines 
symbolize the quantum reductions we have been able to perform, the double 
lines 
symbolize secondary reductions that gives rise to linearizations, and the 
dashed line the classical secondary reduction that is not quantized by our 
method: \\
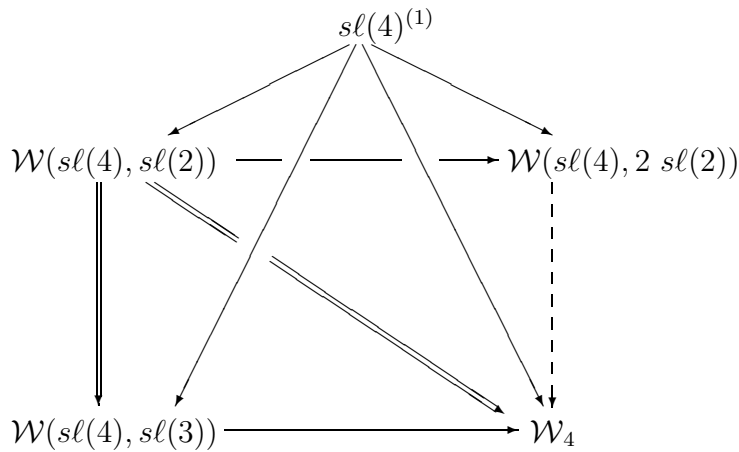
\begin{picture}(540,210)(-210,-90)
\setlength{\unitlength}{0.3mm}
\put(35,100){$s\ell(4)^{(1)}$}
\put(40,96){\vector(-2,-1){80}}
\put(50,96){\vector(2,-1){80}}
\put(43,96){\vector(-1,-2){80}}
\put(46,96){\vector(1,-2){80}}
\put(-110,40) {$\cw(s\ell(4),s\ell(2))$}
\put(110,40) {$\cw(s\ell(4),2~s\ell(2))$}
\put(-110,-80){$\cw(s\ell(4),s\ell(3))$}
\put(120,-80){$\cw_4$}
\put(-10,45) {\line(1,0){20}}
\put(23,45) {\line(1,0){40}}
\put(80,45) {\vector(1,0){26}}
\put(-70,35) {\line(0,-1){95}}
\put(-72,35) {\line(0,-1){95}}
\put(-71,-63) {\vector(0,-1){1}}
\put(-50,35) {\line(3,-2){40}}
\put(-46,35) {\line(3,-2){37}}
\put(5,0) {\line(3,-2){100}}
\put(6,2) {\line(3,-2){100}}
\put(108,-67) {\vector(3,-2){1}}
\put(-15,-75) {\vector(1,0){130}}
\put(130,35){\line(0,-1){5}}
\put(130,25){\line(0,-1){5}}
\put(130,15){\line(0,-1){5}}
\put(130,5){\line(0,-1){5}}
\put(130,-5){\line(0,-1){5}}
\put(130,-15){\line(0,-1){5}}
\put(130,-25){\line(0,-1){5}}
\put(130,-35){\line(0,-1){5}}
\put(130,-45){\line(0,-1){5}}
\put(130,-55){\vector(0,-1){10}}
\end{picture}
 
There are two important consequences that follows from the secondary quantum 
hamiltonian reduction. One of these is the secondary quantum Miura 
transformation. The usual quantum Miura transformation can be used 
to find free field realizations of the \cw~algebras, and in a similar way the 
seondary quantum Miura transformation can be used to find realizations of 
\cw~algebras in terms of subalgebras of other \cw~algebras. For example, in 
the diagram above there are 4 (5 if the dashed line is included) possible 
secondary reductions , and the 
secondary quantum Miura transformation corresponding to these gives us 
realizations of $\cw_4$ in terms of $\cw(s\ell(4),s\ell(2))$ or 
$\cw(s\ell(4),s\ell(3))$, and of    
$\cw(s\ell(4),s\ell(3))$ and $\cw(s\ell(4),2~s\ell(2))$ in terms of 
$\cw(s\ell(4),s\ell(2))$ (and $\cw_4$ in terms of $\cw(s\ell(4), 2\ 
s\ell(2))$ if the dashed line is included). 

The other consequence that follows from the seondary quantum hamiltonian 
reduction is the linearization of \cw~algebras. For a large class of algebras 
$\cw(\cg,\oplus_{n=1}^l \ch_n)$, where the possible $\ch_n$'s are 
given in section \ref{sec:lin}, we can find a secondary hamiltonian reduction 
and a corresponding extended algebra 
$\cw(\cg,\oplus_{n=1}^\ell \ch_n)_{ext}$ which 
is equivalent to a linear algebra with new genrators of positive spin. 
In particular, we are 
able to linearize the \cw$A_n$, 
\cw$C_n$ and \cw$BC_2$ algebras. To take once again the diagram above as 
example, this procedure can give us linearizations of $\cw_4$ and 
$\cw(s\ell(4),s\ell(3))$. This linearization of \cw~algebras could be very 
useful in study of the representation theory of \cw~algebras. In fact one 
could use the linearization to reduce the representation theory of the 
non-linear \cw~algebras to the representation theory of the corresponding 
linear algebras. 

Note that, as mentioned above, we have not in this paper 
exhausted the possible secondary reductions; a number of classical secondary 
reductions cannot be quantized using the present methods, and it would be 
of interest to find a method to quantize these remaining secondary reductions. 

Besides these problems, there are other open questions about the secondary 
quantum hamiltonian reduction. It would be interesting to generalize the 
procedure to supersymmetric \cw~algebras, and to do the secondary quantum 
Miura transformation and the linearization also in that case \cite{MaRa2}. 
Another 
interesting possibility is to study in more detail the linearization of 
\cw~algebras, and to what extent we can actually reduce the 
analysis of the non-linear \cw~algebras to the analysis of the matching linear 
algebra. \\[6pt] 

{\Large {\bf Acknowledgements:}} The authors would like to thank R. Stora for 
stimulating 
discussions. One of the authors (JOM) would like to thank the Niels Bohr 
Institute, where this work was started, and the Danish Natural Science 
Research Council for financial support. 
Finally the authors would like to thank the referee for a very thorough 
referee report, and for making valuable suggestions for improvements of the 
paper. \\[10pt]

\noindent
{\Large \bf Appendices}
\appendix

\sect{Spectral Sequences} 
\label{ap:a}

In this appendix, we will give a few key definitions that are used in 
the theory of spectral sequences. 
For a good introduction to the theory of spectral sequences 
see e.g. \cite{Mc}. 

We assume that we have a complex $(\OM,s)$, i.e. a graded space 
$\OM = \sum_n \OM^n$ and a nilpotent derivation 
$s:\,\OM^n \rightarrow \OM^{n+1}$. We assume furthermore that it is possible to
define a {\em filtration} on the space, i.e. a sequence of subspaces 
$F^q \Omega$ such that 
$$
\{ 0 \} \subset \cdots \subset F^{q+1} \OM \subset F^q\OM \subset 
F^{q-1} \OM \subset \cdots \subset \OM. 
$$
We define a sequence of ``generalized co-cycles" $Z_r^{p,q}$ by 
\bea
Z_r^{p,q} & = & F^q \OM^{p+q} \cap s^{-1} ( F^{q+r} \OM^{p+q+1} ) \nn
& = & \{ x \in F^q \OM^{p+q} | s(x) \in F^{q+r} \OM^{p+q+1} \} 
\eea
We note that it is natural to define 
$$
Z_{\infty}^{p,q} = F^q \OM^{p+q} \cap \ke s. 
$$
We also define a sequence of ``generalized co-boundaries" $B_r^{p,q}$ by 
\bea 
B_r^{p,q} & = & F^q \OM^{p+q} \cap s ( F^{q-r} \OM^{p+q-1} ) \nn
B_\infty^{p,q} & = & F^q \OM^{p+q} \cap \im s, 
\eea
and we see that (suppressing the $(p,q)$ indices) 
$$
\cdots \subset B_{r} \subset B_{r+1} \subset \cdots 
\subset B_\infty \subset Z_\infty \subset \cdots \subset Z_{r+1} \subset 
Z_{r} \subset \cdots 
$$
From these generalized cocycles and coboundaries, we can now define a sequence
of ``generalized cohomologies" $E_r^{p,q}$ by 
\bea
E_0^{p,q} & = & F^q \OM^{p+q} / F^{q+1} \OM^{p+q} \nn
E_r^{p,q} & = & Z_r^{p,q}/ \left ( Z_{r-1}^{p-1,q+1} + B_{r-1}^{p,q} \right ) 
\nn
E_\infty^{p,q} & = & Z_\infty^{p,q}/ \left ( Z_{\infty }^{p-1,q+1} + 
B_{\infty}^{p,q} \right ) 
\eea
It is now possible to show that for every space $E_r$, we can define 
a nilpotent derivative $s_r$. $s_r$ is defined by the commutative diagram: 
\ben
\label{di:0}
\begin{array}{ccc} 
              &  ~~~~~ s ~~~~~  &                 \\ 
Z^{p,q}_r & \longrightarrow & Z^{p+1-r,q+r}_r \\ \\
\eta \downarrow & & \downarrow \eta           \\ \\
E_r^{p,q} & \longrightarrow & E_r^{p+1-r,q+r} \\ 
          &        s_r      & 
\end{array} 
\een
where $\eta$ is the canonical projection operator from $Z_r$ onto $E_r$. 
In other words, for $[x] \in E_r^{p,q}$, $s_r([x]) = [s(x)]$. 

With all these definitions, we are now finally in a position to state the main
theorems of the theory of spectral sequences: \\
The ``generalized cohomologies" $E_r$ that we have introduced are in fact 
cohomologies, namely
\ben 
\label{eqa:th1}
E_{r+1}^{p,q} \cong H^{p,q}( E_r ; s_r)
\een 
If the filtration exhausts all of the space $\Omega$, and if the generalized 
co-cycles $Z^{p,q}_r$ converges to $Z^{p,q}_\infty$, i.e. if 
$\OM = \cup_n F^n \OM$ and $Z_\infty^{p,q} = \cap_r Z_r^{p,q}$, then 
\ben
\label{eqa:th2}
E_\infty^{p,q} \cong F^q H^{p+q} / F^{q+1} H^{p+q}
\een
where $F^q H$ is the filtration on the cohomology $H(\OM;s)$ induced by the
filtration on $\OM$: 
$$
F^q H^{p+q} = H^{p+q} (F^q\OM;s)
$$
This is the principal result of the theory of spectral sequences. 
It gives us a way 
to find the cohomology $H(\OM;s)$, supposing that we are able to use the
knowledge of the spaces $F^q H^{p+q} / F^{q+1} H^{p+q}$ to reconstruct 
$H(\OM;s)$. The usefulness of the spectral sequences rests on the fact that 
in practical application the spectral sequence often collapses after a few 
steps, i.e. $s_r$ is identically zero for $r > r_0$ where
$r_0$ is some low number. 

Let us show the following 
\begin{lem}
\label{eq:recH}
If $F^q \OM = 0$ for $q>0$, then 
$H(\OM;s) \cong E_\infty$
\end{lem}
Namely $F^q \OM = 0$ for $q>0$ implies that 
$F^q H^{p+q} = 0$ for $q>0$. This means that
\bea
E_\infty^{p,0} & \cong & F^0 H^{p} / F^{1} H^{p} \cong F^0 H^p \\
E_\infty^{p+1,-1} & \cong & F^{-1} H^p / F^0 H^p 
\eea
etc..., and we can use this to show that 
$$
F^{-r} H^p \cong E_\infty^{p+r,-r} \oplus \cdots \oplus E_\infty^{p,0} 
$$
or equivalently (since $E_\infty^{p,q} = 0$ for $q>0$)
$$
H^p(\OM;s) \cong \sum_{r\in\z} E_\infty^{p+r,-r} 
$$
which proves the lemma \ref{eq:recH}.

Let us mention here that the isomorphism 
$E_\infty^{p,q} \cong F^q H^{p+q} / F^{q+1} H^{p+q}$ is in general a
vector space isomorphism. 
If the space $\OM$ in addition to being a vector space is
also an algebra (as in the case that we are interested in here) then 
if we can define algebras on all the cohomologies in the spectral
sequence such that the algebra on 
$E_0^{p,q} = F^q \OM^{p+q} / F^{q+1} \OM^{p+q}$ is induced by 
the algebra on $F^q \OM^{p+q}$, i.e. 
$[a], [b] \in E_0: [a] \circ [b] = [a\circ b]$ (where $\circ$ denotes the
algebra composition), and the algebra on $E_{r+1} = H(E_r,s_r)$ is induced by 
the algebra on $E_r$, then the the isomorphism 
$E_\infty^{p,q} \cong F^q H^{p+q}/F^{q+1} H^{p+q}$ 
is an algebra isomorphism. However, even if  
$E_\infty^{p,q} \cong F^q H^{p+q}/F^{q+1} H^{p+q}$ 
is an algebra isomorphism, it may be nontrivial to reconstruct 
the algebra of $H^n(\OM;s)$. 

In the case where the complex $(\OM;s)$ can be given the structure of a 
double complex structure $(\OM;s',s'')$ with two anti-commuting nilpotent 
operators $s'$ and $s''$, and with a bigrading $\OM = \sum_{p,q} \OM^{p,q}$, 
the spectral sequence simplifies somewhat. Define $s=s'+s''$.
The filtration is defined in terms 
of the bi-grading as 
\bea
F^q \OM & = & \bigoplus_{i\in\z,j\geq q} \OM^{i,j} \nn
F^q \OM^{p+q} & = & \bigoplus_{i\geq 0} \OM^{p-i,q+i} 
\eea

The first element in the spectral sequence, $E_0$, is defined by 
\ben
\label{eq:e0}
E_0^{p,q} = F^q \OM^{p+q} / F^{q+1} \OM^{p+q} 
\cong \OM^{p,q},
\een
and the nilpotent bigrade $(1,0)$-operator $s_0$ on $E_0$ 
is defined by the commutative diagram 
\ben
\label{di:1}
\begin{array}{ccc} 
          &  ~~~~~ s ~~~~~  &                 \\ 
F^q \OM^{p+q} & \longrightarrow & F^q \OM^{p+q+1} \\ \\
\eta \downarrow & & \downarrow \eta           \\ \\
E_0^{p,q} & \longrightarrow & E_0^{p+1,q} \\ 
          &        s_0      & 
\end{array} 
\een
where $\eta$ is the canonical projection operator 
$$
F^q \OM^{p+q} \ab{\eta}{\rightarrow} 
F^q \OM^{p+q} / F^{q+1} \OM^{p+q} \cong \OM^{p,q}. 
$$ 
This means that if we identify $x\in E_0^{p,q}$ with $x \in F^q \OM^{p+q}$,
then $s_0 (x) = \eta(s(x))$ -- and since $\eta$ here is the projection operator
on $\OM^{p+1,q-1}$, $s_0$ is simply the bigrade $(1,0)$ part of $s$: 
$s_0 = s'$. 

The second element in the spectral sequence is 
\bea
E_1^{p,q} & = & Z_1^{p,q} / (Z_0^{p-1,q+1} + B_0^{p,q}) \nn 
& \cong & \frac
{\OM^{p,q} \cap s^{-1}( \OM^{p,q+1}) } 
{\OM^{p,q} \cap s(\OM^{p-1,q})} 
\eea
Note that 
$\OM^{p,q} \cap s^{-1}( \OM^{p,q+1}) = \OM^{p,q} \cap \ke s_0$ and 
$\OM^{p,q} \cap s(\OM^{p-1,q}) = \OM^{p,q} \cap \im s_0$; so, in
agreement with equation (\ref{eqa:th1}), we can also write $E_1$ as 
\ben 
\label{eqa:coho}
E_1^{p,q} = H^{p,q} (E_0; s_0)
\een
The operator $s_1$ on $E_1$ is again defined by the commutative diagram 
\ben
\label{di:2}
\begin{array}{ccc} 
          &  ~~~~~ s ~~~~~  &                 \\ 
Z_1^{p,q} & \longrightarrow & Z_1^{p,q+1} \\ \\
\eta \downarrow & & \downarrow \eta           \\ \\
E_1^{p,q} & \longrightarrow & E_1^{p,q+1} \\ 
          &        s_1      & 
\end{array} 
\een
and $\eta$ is again the canonical projection operator. Consider 
$x\in \ke s_0$ and let $[x] = \eta(x)$ be the corresponding equivalence class
in $E_1$. Then $s_1([x]) = \eta(s(x))$ is just the bigrade $(0,1)$-part of 
$s(x)$, projected on $E_1$, i.e.:
$$
s_1([x]) = [s'(x)].   
$$

\sect{Shift of the Constraints Using a $U(1)$ Generator}
\label{sec:u1}

\indent

We are looking for couples of $\cw(\cg,\ch)$ algebras such that the sets of 
first class constraints 
are embedded one into the other. We first consider the case $\cg=s\ell(N)$, 
and to clarify 
the presentation, we focus on the secondary reductions of type 
$\cw(\cg,\ch)\ \rightarrow\ 
\cw(\cg,\cg)\equiv\cw(\cg)$.

\indent

In $s\ell(N)$, the regular subalgebras \ch\ can always be chosen in such 
a way that the 
simple roots of  $\ch$ are also simple roots of $\cg$. Let 
$\ch=\oplus_{n=1}^\ell \ch_n$ 
where $\ch_n$ 
are simple subalgebras of rank $r_n=\mbox{rank}(\ch_n)$, ordered in such a 
way that
$r_n\leq r_m$ if $n>m$. We define as simple roots 
\be
\begin{array}{ll}
\mbox{Simple roots of }\cg & \alpha_1,...,\alpha_{r_1}; \alpha_{\rho_1};  
\alpha_{\rho_1+1},..., \alpha_{\rho_1+r_2}; \alpha_{\rho_2};  
\alpha_{\rho_2+1},..., \alpha_{\rho_2+r_3};\alpha_{\rho_3}; ...\\
 & ...;\alpha_{\rho_{\ell-1}}; \alpha_{\rho_{\ell-1}+1},..., 
\alpha_{\rho_{\ell-1}+r_{\ell}};\alpha_{\rho_{\ell}+1},...,\alpha_{N-1}\\ 
\mbox{Simple roots of }\ch & \alpha_1,...,\alpha_{r_1};
\phantom{\alpha_{\rho_1}}; 
\alpha_{\rho_1+1},..., \alpha_{\rho_1+r_2};\phantom{\alpha_{\rho_2}};  
\alpha_{\rho_2+1},..., \alpha_{\rho_2+r_3};\phantom{\alpha_{\rho_3}}; ...\\
 & ...; \phantom{\alpha_{\rho_{\ell-1}}}; \alpha_{\rho_{\ell-1}+1},..., 
\alpha_{\rho_{\ell-1}+r_{\ell}}\\ 
\mbox{ with  } & \rho_n=\sum_{i=1}^n (r_i+1)
\end{array}
\ee
In the fundamental representation of $s\ell(N)$, this simply means that we 
have divided the 
$N\times N$ matrix into $r_j\times r_j$ blocks of decreasing size, plus 
(when it exists) a block
$(N-\rho_\ell)\times (N-\rho_\ell)$.

The gradation associated to the Cartan generator of the principal $s\ell(2)$ 
in $\ch$ 
attributes a grade 1 to each simple root of $\ch$, but the grade of the 
simple roots of type 
$\alpha_{\rho_n}$ is 
\be
gr(\alpha_{\rho_n})=-\frac{r_n+r_{n+1}}{2}<0 
\ee
where we have set $r_{\ell+1}=0$.
This implies that the root generators $E_{\rho_n}$ are constrained in 
$\cw(\cg, \ch)$ although they are not in $\cw(\cg)$.
Thus, it is clear that we have to introduce a new gradation such that 
\be
gr(\alpha_{\rho_n})'\geq0 \label{gr>}
\ee
while not changing the resulting $\cw$-algebra. Let $H$ be the gradation 
we are looking 
for. Then, if $M_0$ is the Cartan generator of the $s\ell(2)$ embedding we 
are considering 
(it has not been changed because we want the $\cw$-algebra to be the same), 
the new 
gradation is characterized by the generator $U=H-M_0$ which commutes with the 
$s\ell(2)$ algebra and which ''respects" the highest weight gauge. Thus, 
classifying the 
different gradations $H$ is the same as classifying the different $U(1)$ 
generators 
submitted to the non-degeneracy condition
\be
\mbox{ker }ad(M_+)\cap \cg'_-={0} \label{nonDeg}
\ee
where $\cg'_-$ denotes the subalgebra of $\cg$-generators which have 
negative grade 
w.r.t. $H$.
This technique has been developed in \cite{FrRaSo,U1}, where all the possible 
gradations leading to the same $\cw$-algebra have been classified. 
The procedure goes 
along the following lines. 

We start with the decomposition  of the fundamental of $s\ell(N)$ w.r.t. 
the principal 
$s\ell(2)$ in $\ch$:
\be
\underline{N}=\oplus_{\mu=1}^I n_\mu\ D_{j_\mu} \mb{with} j_\mu\neq j_\nu 
\mbox{ when } \mu \neq \nu
\ee
and add the following $U(1)$ eigenvalues
\be
\underline{N}=\oplus_{\mu=1}^I n_\mu\ D_{j_\mu}(y_\mu) 
\ee
Then, computing the adjoint representation from this decomposition of the 
fundamental, 
\be
\cg= \oplus_{k} D_{k}(Y_k) 
\ee
where the $Y_k$'s are differences of two $y_\mu$'s. The eigenvalues of the 
allowed $U(1)$ generators will be characterized by the equations
\be
|Y_k| \leq k \mb{and} Y_k \in \half\ \Z\ ,\ \ \forall\ D_{k}(Y_k) 
\label{nonDeg2}
\ee
Then, the different gradations will 
be $M_0+U$, with $M_0$ the Cartan generator of the $s\ell(2)$ under 
consideration, and 
$U$ one of the allowed $U(1)$ generators.

Now, to get a gradation satisfying both equations (\ref{gr>}) and 
(\ref{nonDeg2}),  we have to impose
\be
|y_\mu-y_\nu| \leq |j_\mu-j_\nu| \mb{and} y_\mu-y_\nu\geq j_\mu+j_\nu
\ee
which is clearly satisfied only if one of the two $j$'s is zero, ie if 
$\ch$ is 
simple\footnote{If $\ch$ is simple, there will be 
only one $D_j$ representation with $j\neq0$ in the fundamental of $\cg$}. 
In that case, 
the $s\ell(2)\oplus U(1)$ decomposition
\be
D_j(y)\oplus (N-2j-1)D_0(z) \mb{with} y=\frac{j(N-2j-1)}{N} \mb{and} z=-
\frac{j(2j+1)}{N}
\ee
indeed gives a gradation where all the simple roots of $s\ell(N)$ have 
positive grades, and 
whose associated $\cw$-algebra is $\cw(s\ell(N),\ch)$.

For the general secondary reduction 
$\cw(s\ell(N),\ch')\ \rightarrow\ \cw(s\ell(N),\ch)$, the 
reasoning follows along the same lines. We however have to look at the 
grade of all the roots (since some simple roots have negative grades in 
the general case). Then, one asks the gradations to satisfy 
$\cg_-\subset\cg_-'$. This necessary condition is sufficent in the 
case of $s\ell(N)$ because the simple roots of $\ch$ can always being 
choosen among the simple roots of $\ch'$ (and thus the constraints 
$J^{\alpha_i}=1$ for \ch\ are a subset of the constraints 
$J^{\alpha_i}=1$ for $\ch'$). After a tedious calculation, and 
using non-degenerated $U(1)$ generators both for \ch\ and $\ch'$, 
one gets the following property:

\begin{prop}\label{U1-sl}

In the case of secondary reductions of type 
$\cw(s\ell(N),\ch')\ \rightarrow\ \cw(s\ell(N),\ch)$, we have the following 
necessary and sufficient condition for the existence of a $U(1)$ generator 
which satisfies both the non-degeneracy (\ref{nonDeg}) and the embbeding
 of the set of constraints associated to \ch\ into the set of constraints 
associated to $\ch'$:

If $\ch'$ decomposes as 
\be
\ch'=\oplus_{\alpha=1}^l m_\alpha\ s\ell(p_\alpha)
\ee
the $U(1)$ generator exists iff $\ch$ decomposes as:
\be
\ch=\oplus_{\alpha=1}^l m_\alpha\ s\ell(q_\alpha) \mb{with} \left\{
\begin{array}{l} 
2\leq p_\alpha \leq q_\alpha \ \ \ \forall\alpha \\
| p_\alpha- p_\beta| \leq |q_\alpha- q_\beta|  \ \ \ \forall\alpha, \beta 
\end{array}\right.
\ee
\end{prop}

\indent

Now, turning to the case of orthogonal and symplectic algebras, we can do 
the same 
calculation. However, for these algebras, the $U(1)$ generator is much 
more constrained 
(see \cite{FrRaSo}, sections 5.2 and 5.3) so that there are less $U(1)$ 
generators 
satisfying both the non-degeneracy and the embedding conditions. Note 
that one has really to 
check in each case that the sets of currents constrained to 1 are also 
embedded one into the 
other, since the simple roots of $\ch'$ are not always simple roots of \ch.

Apart from these restrictions, the calculation is the same as for 
$s\ell(N)$ algebras, so that one is led to

\begin{prop}\label{U1-sosp}

In the case of secondary reductions of type 
$\cw(\cg,\ch')\ \rightarrow\ \cw(\cg,\ch)$ with $\cg=so(N)$ or $sp(N)$, we 
have the following necessary and sufficient condition for the existence of 
a $U(1)$ generator which satisfies both the non-degeneracy (\ref{nonDeg}) 
and the 
embedding of the sets of constraints.

\indent
$-$ For $so(N)$, $\ch$ and $\ch'$ must be of the form 
\[
\left\{ \begin{array}{l} \ch'= (n+1)\ so(p) \\
 \ch= n\ so(p)\ \oplus\ so(p+2)\end{array} \right. \mb{with} 
\left\{ \begin{array}{l} N= (n+1) p+2\ ;\ n\geq0 \\ 
N\equiv p \ \ [\mbox{mod }2] \end{array} \right.
\] 

\indent $-$ For $sp(N)$, $\ch$ and $\ch'$ must be of the form 
\[
\left\{ \begin{array}{l}\ch'= s\ell(2) \oplus_\mu s\ell(2p_\mu) \\
 \ch= sp(4) \oplus_\mu s\ell(2p_\mu) \end{array} \right.
 \mb{with} p_\mu\in \mbox{\N}
\] 
or of the form
\[
\left\{ \begin{array}{l}\ch'= s\ell(2)\\
\ch=\oplus_j sp(2q_j) \oplus_\mu s\ell(p_\mu) \end{array} \right.
 \mbox{ with } 
 \left\{ \begin{array}{l}
 \mbox{{\rm either }}p_1\in 2\N, p_1\geq p_\mu+1\ \forall \mu\geq2 \mbox{ and }
 p_1\geq 2q_j+1\ \forall j \\
 \mbox{{\rm or }}p_1\in (2\N+1), p_1\geq p_\mu+2\ \forall \mu\geq2 \mbox{ and }
 p_1\geq 2q_j+1\ \forall j \\
 \mbox{{\rm or }}q_1\geq q_j+1\ \forall j\geq2 \mbox{ and }
 q_1\geq \half(p_\mu+1)\ \forall \mu 
  \end{array} \right.
\] 
\end{prop}

Let us remark that in the case $\cg=so(5)$, the $U(1)$ generator exists when
considering  
the reduction $\cw(so(5),so(3))\ \rightarrow\ \cw(so(5))$, while in the case 
$\cg=sp(4)$ the $U(1)$ generator exists for the reduction 
$\cw(sp(4), s\ell(2))\ \rightarrow\ \cw(sp(4))$ which is in agreement with the 
isomorphism between the $so(5)$ and $sp(4)$ algebras.

\end{document}